\documentclass[aps, pre, reprint, superscriptaddress]{revtex4-2}
\usepackage{amsmath,amssymb}
\usepackage{bm}
\usepackage{mathptmx}
\usepackage{graphicx}
\usepackage{color}
\usepackage{hyperref}
\usepackage{float}
\usepackage{orcidlink}
\usepackage{microtype}
\usepackage{xspace}
\usepackage{makecell}
\usepackage{nicefrac}
\usepackage[normalem]{ulem}
\usepackage{tikz}
\usetikzlibrary{arrows.meta}
\usepackage{wasysym}
\usetikzlibrary{positioning}
\graphicspath{{./Figures}}

\newcommand{\T}[1]{\textcolor{black}{#1}} 

\begin{document}

\title{The limits of higher-order interactions in shaping chimera states}
 \affiliation{Department of Physics, FFCLRP, University of S\~ao Paulo, 14040-900 Ribeir\~ao Preto, SP, Brazil}

 \author{Ediline L.F.~Nguessap\,\orcidlink{0009-0002-2854-9794}}
 \affiliation{Department of Physics, FFCLRP, University of S\~ao Paulo, 14040-900 Ribeir\~ao Preto, SP, Brazil}

 \author{Thierry Njougouo\,\orcidlink{0000-0001-7706-7674}}
\email[Contact author: ]{thierry.njougouo@imtlucca.it}
\affiliation{IMT School for Advanced Studies,
Piazza San Francesco 19, 55100 Lucca, Italy}

 \author{Fernando Fagundes Ferreira\,\orcidlink{0000-0001-9183-8287}}
 \affiliation{Department of Physics, FFCLRP, University of S\~ao Paulo, 14040-900 Ribeir\~ao Preto, SP, Brazil}

 \begin{abstract}
In neuronal networks, chimera states, characterized by the coexistence of coherent and incoherent populations, have attracted considerable attention in the study of collective brain dynamics, due to its potential implications for neurological disorders such as epilepsy. Their emergence has been extensively investigated with a particular emphasis on the role of network topology, both in conventional pairwise interactions and, more recently, in higher-order interaction networks. Here, we propose a generalized framework for neuronal hypergraphs incorporating higher-order interactions through attractive chemical and repulsive electrical coupling. Using networks of FitzHugh--Nagumo neurons with interaction orders $d=2,3,4$, corresponding to
$(d+1)$-body interactions within a framework that can be extended to higher $d$, we numerically investigate the emergence of chimera states under the combined effect of attractive and repulsive coupling and we compare the resulting dynamics with those of their corresponding clique-projected networks. Our results reveal that, both dynamical similarities and qualitative differences between genuine higher-order interactions on hypergraphs and their pairwise representations. While higher-order interactions often support broader chimera regions, we show that clique projections can also preserve or promote chimera states, even in regimes where the corresponding hypergraphs exhibit fully incoherent dynamics. Therefore, these findings suggest that chimera formation is  not only influenced by the topology or by the interaction order but also by the form of the coupling functions and the balance between attractive and repulsive interactions, offering further insights into collective dynamics in neuronal networks.
 \end{abstract}

\maketitle

\section{Introduction}
In complex systems, chimera states constitute a paradigmatic manifestation of spontaneous symmetry breaking, where homogeneous populations of identically coupled oscillators spontaneously self-organize into coexisting coherent and incoherent subpopulations. This self-organization reveals how structurally symmetric interactions can give rise to heterogeneous collective dynamics and emergent spatiotemporal coexistence of coherent and incoherent oscillators. First reported in nonlocally coupled phase oscillators by Kuramoto and Battogtokh~\cite{kuramoto2002coexistence} and subsequently named \textit{chimera states} by Abrams and Strogatz~\cite{abrams2004chimera}, they have since been experimentally observed in diverse physical systems, including chemical and mechanical systems~\cite{tinsley2012chimera,martens2013chimera}. In neuroscience, chimera states have attracted particular interest because they provide a natural mechanism for localized, partially coherent activity. Prominent examples include unihemispheric slow-wave sleep~\cite{rattenborg2007non}, and compelling analogies have been drawn between the sudden collapse of chimera states and the onset of epileptic seizures~\cite{andrzejak2016all}.

This biological motivation has led to extensive investigations of chimera states in neuronal networks, primarily using excitable and spiking neuron models, such as the FitzHugh--Nagumo (FHN) and Hindmarsh--Rose systems~\cite{fitzhugh1961impulses,omelchenko2013nonlocal,
calim2020chimera,majhi2017chimera,simo2021chimera,hussain2021synchronization,
nguessap2026dynamic}. Unlike phase oscillators, these models explicitly capture spike generation and nonlinear excitability. \T{Furthermore, biological neurons communicate through two fundamentally distinct synaptic mechanisms, named \textit{electrical} and \textit{chemical} synapses. The electrical synapses, mediated by gap junctions, primarily support diffusive coupling, and chemical synapses, involve nonlinear, voltage-dependent transmission governed by pre-synaptic activity and postsynaptic receptor
dynamics~\cite{izhikevich2006dynamical,gerstner2002spiking}. The interplay between these two coupling mechanisms, specifically the competition between attractive and repulsive interactions, plays a crucial role in shaping the emergence and the stability of chimera
states~\cite{belykh2019synchronization,calim2020chimera,
simo2021chimera,jiang2020chimera,paul2024attractive}.}  Such competition can be viewed as a simplified representation of the balance between excitatory and inhibitory influences in neural circuits~\cite{bennett2004electrical,mirzaei2023synchronization}. However, a fundamental limitation of these existing studies is that they assume interactions occur exclusively between pairs of units~\cite{newman2003structure}.

Many natural systems, including neuronal networks, involve cooperative interactions among groups of three or more elements that cannot be fully represented as a superposition of independent pairwise interactions~\cite{battiston2020networks,bianconi2021higher,battiston2021physics}. In the brain, dendritic integration and the coordinated activity of neuronal assemblies give rise to intrinsically many-body effects~\cite{petri2014homological,sizemore2018cliques,poirazi2003pyramidal,london2005dendritic,major2013active}. Hypergraphs and simplicial complexes provide a natural mathematical framework to capture these genuine higher-order interactions (HOIs)~\cite{bick2023higher,boccaletti2023structure,majhi2022dynamics}. Theoretical studies have demonstrated that, these HOIs can generate dynamical regimes inaccessible to pairwise networks, such as abrupt synchronization transitions, extensive multistability, and complex cluster states~\cite{tanaka2011multistable,lucas2020multiorder,skardal2019abrupt,millan2020explosive,gambuzza2021stability,zhang2023higher}. For chimera states in particular, HOIs have been shown to promote their emergence, enlarge their parameter regions, or fundamentally modify their structure in phase-oscillator and other low-dimensional models \cite{kundu2022higher,majhi2022dynamics,ghosh2024chimeric,muolo2024phase,ji2024chimera,muolo2025pinning}. However, in the context of biologically motivated neuronal networks, the impact of HOIs has been investigated largely in relation to global synchronization and explosive transitions ~\cite{parastesh2022synchronization,ghosh2024chimeric,
singh2024higher,farrera2025explosive}. Consequently, this raises a fundamental research question: To what extent does the interplay between genuine higher-order electrical and chemical interactions govern the emergence and stability of chimera states in excitable neuronal networks?

Beyond this dynamical perspective, a broader methodological issue concerns the extent to which genuine HOIs can be accurately represented by pairwise approximations. A common used approach is clique projection, in which each hyperedge of the network is replaced by a fully connected set of nodes interacting trougth pairwise links. Although this transformation preserves pairwise connectivity, it generally discards information about the original many-body interaction structure. Moreover, when higher-order coupling functions are replaced by its corresponding pairwise contributions, the resulting collective dynamics are not necessarily preserved. Previous work indicates that chimera states can be weakened or suppressed after such a projection \cite{kundu2022higher,muolo2024phase,muolo2025pinning}. It is not clear whether this holds generally or whether it depends on the form of the coupling functions and on the balance between attractive and repulsive interactions. This question is related to the broader problem of when higher-order interactions can be reduced to pairwise ones \cite{llabres2026reducibility}.

\T{This work addresses both aspects by introducing an hypergraph framework for FHN neurons, which can be extended to other neuronal networks, incorporating competing repulsive electrical and attractive chemical higher-order coupling. We consider interaction orders $d=2,3,4$, corresponding to three-, four-, and five-body interactions, or hyper-edge sizes, respectively, on nonlocal, and nonregular hypergraph structures as illustrated in~\cite{muolo2024phase}. Based on the strength of incoherence proposed by Gopal et al.~\cite{gopal2014observation}, and spatiotemporal patterns, we explore the emergence of coherent, incoherent, and chimera states across the two-dimensional coupling-strength parameter space, then, comparing the resulting dynamics with those of the corresponding clique-projected networks. Our results reveal that, although genuine HOIs often favor broader regions of chimera existence, their clique-projected counterparts can also preserves or even promote chimera states in parameter spaces where the original hypergraphs remain fully incoherent. Therefore, our findings suggest that chimera states is governed not solely by network topology or group sizes, but also by the functional structure of the coupling mechanisms and the competition between attractive and repulsive interactions.} The remainder of this paper is organized as follows. Sec.~\ref{sec:model} details the mathematical framework, introducing the higher-order neuronal hypergraph model and its pairwise counterpart obtained via clique projection. Sec.~\ref{sec:Results} presents the numerical results, and systematically compares the collective dynamics emerging under attractive-repulsive coupling in both network representations. Finally, Sec.~\ref{sec:conclusion} summarizes the main findings and draws the conclusions of this work.  

\section{Model} \label{sec:model}
\subsection{Higher-Order Interactions in Pure Hypergraphs} \label{ssec:HOIs}

We consider a network of $N$ identical FHN oscillators, whose isolated dynamics is governed by
\begin{equation}
\begin{aligned}
\epsilon \dot{x} &= x -\frac{x^3}{3} - y,\\
\dot{y} &= x + a,
\end{aligned}
\label{eq:FHNiso}
\end{equation}
where $x$ and $y$ are the activator and inhibitor variables, respectively. The parameter $\epsilon=0.05$ determines the time-scale separation, while $a=0.5$ controls the excitability threshold, placing each isolated oscillator in the excitable regime~\cite{fitzhugh1961impulses}.
The oscillators are coupled through higher-order interactions represented by a $d$-uniform, undirected hypergraph, in which each hyperedge simultaneously connects $(d+1)$ oscillators. We consider the same nonlocal and nonregular hypergraph topology introduced in Ref.~\cite{muolo2024phase}, which was shown to enhance chimera states. To account for neuronal communication, we extend both electrical and chemical synaptic coupling functions to the hypergraph framework, thereby constructing a higher-order network model with many-body electrical and chemical interactions.

In the hypergraph representation, let $\mathcal{E}_i$ denote the set of hyperedges incident to node $i$. Since an oscillator may belong to several overlapping hyperedges, its dynamics results from the collective interactions within all hyperedges containing it, rather than from a superposition of independent pairwise interactions. This hypergraph representation naturally captures genuine many-body interactions among neighboring oscillators. Accordingly, within this general higher-order interaction framework, the dynamics of the $i$th oscillator is governed by Eq.~\eqref{eq:FHNhoi}
\begin{equation}
\begin{aligned}
\varepsilon \dot{x}_i
&=
x_i-\frac{x_i^3}{3}-y_i
+{\sigma_e}
\sum_{j_1,\ldots,j_d=1}^{N}
{A}^{(d)}_{i j_1\cdots j_d}
\left[
\mathcal{P}^{(d)}(x_{j_1},\ldots,x_{j_d})
-x_i^{p_d}
\right]
\\
&\quad
+{\sigma_c}(x_s-x_i)
\sum_{j_1,\ldots,j_d=1}^{N}
{A}^{(d)}_{i j_1\cdots j_d}
\prod_{\ell=1}^{d}S(x_{j_\ell}),
\\
\dot{y}_i
&=
x_i+a,
\end{aligned}
\label{eq:FHNhoi}
\end{equation}
where ${A}^{(d)}_{i j_1\cdots j_d}$ is the symmetric
adjacency tensor of the $(d+1)$-uniform hypergraph, defined by
\[
{A}^{(d)}_{i j_1\cdots j_d}
=
\begin{cases}
1,
& \{i,j_1,\ldots,j_d\}\in\mathcal{E},\\
0,
& \text{otherwise}.
\end{cases}
\]
Following the diffusive-like higher-order interaction framework introduced in Ref.~\cite{muolo2024phase}, the electrical interaction is constructed using nonlinear polynomial functions of odd degree defined in Eq.~\eqref{eq:parity_rule}.
\begin{equation}
\mathcal{P}^{(d)}(x_{j_1},\ldots,x_{j_d}) =
\begin{cases}
    \displaystyle
        \prod_{\ell=1}^{d}x_{j_\ell},& d\ \mathrm{odd}, \\[1mm]
    \displaystyle
        x_{j_1}^{2}\prod_{\substack{\ell=2}}^{d} x_{j_\ell}, & d\ \mathrm{even},
\end{cases}
\label{eq:parity_rule}
\end{equation}
where the exponent or polynomial degree $p_d$ is defined as
\[ \qquad
p_d=
\begin{cases}
d,
& d\ \mathrm{odd},
\\
d+1,
& d\ \mathrm{even}.
\end{cases} \]

One of the most important aspects of this framework is that it ensures that the electrical coupling functions are odd under a simultaneous sign reversal of all their arguments, namely $\mathcal{P}^{(d)}(-x_{j_1},\ldots,-x_{j_d})
= -\mathcal{P}^{(d)}(x_{j_1},\ldots,x_{j_d})$. Moreover, these functions satisfy $\mathcal{P}^{(d)}(x,\ldots,x)=x^{p_d}$, ensuring that the electrical coupling vanishes under complete synchronization. So, for even $d$, our polynomial considered is not individually invariant under permutations of its arguments, since the first variable appears quadratically. Nevertheless, by definition of the adjacency tensor the electrical coupling in the even $d$ is permutation-invariant due to the symmetry of the adjacency tensor and the summation over all ordered index combinations.

Moreover, the higher-order chemical interaction is governed by the
sigmoidal activation function $S(x)$, defined as~\cite{simo2021chimera}
\begin{equation}
S(x)
=
\frac{1}
{1+\exp\left[-\lambda(x-\theta_s)\right]},
\label{eq:sigmoid}
\end{equation}
where $\lambda=10$ controls the steepness of the activation function
and $\theta_s=-0.25$ denotes the synaptic activation threshold. The
reversal potential is fixed at $x_s=2.0$. Throughout this investigation, we focus on a particular coupling configuration, referred to as attractive--repulsive coupling~\cite{jiang2020chimera,paul2024attractive}, in which the higher-order chemical interaction is always attractive, whereas the higher-order electrical interaction is repulsive. Accordingly, $\sigma_c>0$ corresponds to attractive higher-order chemical interactions, while $\sigma_e<0$ denote repulsive higher-order electrical interactions, respectively. Since the FHN model is a paradigmatic model of excitable neurons, these coupling function framework may also be viewed from a neuroscience perspective. In this context, the interplay between the attractive chemical coupling and the repulsive electrical coupling may provide a simplified representation of the balance between excitatory and inhibitory interactions in neuronal networks. While the attractive chemical coupling tends to promote synchronization or coherence by bringing neuronal activities closer together~\cite{bennett2004electrical}, the repulsive electrical coupling counteracts global coherence by enhancing local dynamical differences~\cite{mirzaei2023synchronization}. Such a balance may contribute to the emergence of partially synchronized states, including chimera-like patterns, which have been proposed as a possible mechanisn underlying localized synchronisation or coherence observed in neural systems.

\subsection{From Hypergraph to Pairwise Interactions via Clique Projection} \label{ssec:clique}

In order to explicitly contrast the hypergraph dynamics with its pairwise counterpart, let us construct the clique projection of the original hypergraph model defined in Eq.~\eqref{eq:FHNhoi}. In this hypergraph formulation, the (d+1) nodes of each hyperedge interact simultaneously through a genuine many-body coupling function. However, in the clique-projected network, each hyperedge is replaced by a complete graph, so that the same group of nodes interacts only through pairwise links. This representation preserves the node groups associated with the original hyperedges while replacing collective higher-order interactions by sums of binary interactions [see Ref.\cite{muolo2024phase}]. The corresponding pairwise coupling terms are given by Eq.~\eqref{eq:FHN_projected}
\begin{equation}
\begin{aligned}
\varepsilon \dot{x}_i
&=
x_i-\frac{x_i^3}{3}-y_i
+
\sigma_e
\sum_{j=1}^{N}
A_{ij}^{(P)}
\left(
x_j^{q}-x_i^{q}
\right)
\\
&\quad
+
\sigma_c(x_s-x_i)
\sum_{j=1}^{N}
A_{ij}^{(P)}
\left[S(x_j)\right]^q,
\\
\dot{y}_i
&=
x_i+a,
\end{aligned}
\label{eq:FHN_projected}
\end{equation}
where $A_{ij}^{(P)}$ denotes the adjacency matrix of the clique-projected graph, and the exponent $q$ is chosen to match the effective interaction order of the corresponding hypergraph.
Crucially, modeling genuine HOIs via hypergraphs, rather than reducing them to pairwise clique projections, carries direct biophysical implications. In the hypergraph formulation, the multiplicative chemical coupling requires the simultaneous activation of all neurons within a hyperedge. Biologically, this interaction captures nonlinear dendritic integration and coincidence detection mechanisms such as NMDA receptor activation and local dendritic spikes, which generate powerful supralinearities essential for complex neural computations~\cite{poirazi2003pyramidal,london2005dendritic,major2013active,altshuler2026moving}. By contrast, the clique projection simplifies this collective process into a linear summation of independent pairwise interactions. This corresponds to standard AMPA receptor activation, where postsynaptic responses are constructed via the linear integration of individual synaptic inputs~\cite{polsky2004computational,bicknell2021synaptic,pagkalos2023introducing}. Although both representations preserve the exact same structural connectivity (i.e., the same groups of connected neurons), they encode fundamentally different biophysical interaction rules. This mathematical distinction demonstrates that the functional form of coupling is as important as the structural topology itself in neuronal modeling.

\section{Results} \label{sec:Results}

For the numerical simulations, we consider a network of $N=504$ FHN oscillators interacting on the same underlying topology, modeled either as a $d$-uniform, nonlocal, and non-regular hypergraph supporting higher-order interactions (Sec.~\ref{ssec:HOIs}) or as its clique-projected graph supporting only pairwise interactions (Sec.~\ref{ssec:clique}). The network dynamics is integrated using the classical fourth-order Runge--Kutta method with a fixed time step $dt=0.05$. Each simulation is performed over $T=1500$ time units, corresponding to $3\times10^4$ integration steps. To eliminate transient effects, the first $97\%$  of the simulated trajectory is discarded, and all reported quantities are computed from the remaining asymptotic dynamics.
Let us begin by analyzing the collective dynamics arising from the simultaneous action of repulsive (i.e., $\sigma_e<0$) higher-order electrical interactions, and attractive higher-order chemical interactions (i.e., $\sigma_c>0$). In this investigation, particular attention is devoted to analyzing how these competing interaction mechanisms govern the emergence of coherent, incoherent, and chimera states.

To quantitatively characterize the emergence of chimera states in both hypergraph and pairwise interactions, we employ the \textit{Strength of Incoherence} (SI) introduced in Ref.~\cite{gopal2014observation}, which has become a standard measure for distinguishing coherent, incoherent, and chimera states in coupled dynamical systems. The local amplitude difference is first defined as $z_i=x_i-x_{i+1}$, $i=1,\ldots,N$, with periodic boundary conditions. The same definition can equally be applied to the inhibitor variable $y$; throughout this work, however, we use the activator variable $x$ for convenience. The network is then partitioned into $M$ equal spatial bins, each containing $n=N/M$ oscillators. Therefore, for the $m$th bin, the local standard deviation is computed as
\begin{equation}
\sigma_m=
\left\langle
\sqrt{\frac{1}{n}
\sum_{j=n(m-1)+1}^{nm}
\left(z_j-\langle z_m\rangle\right)^2}
\right\rangle_t,
\end{equation}
where
\begin{equation}
\langle z_m\rangle =
\frac{1}{n}
\sum_{j=n(m-1)+1}^{nm}z_j
\end{equation}
is the local mean within the $m$th bin, and $\langle\cdot\rangle_t$ denotes time averaging over the asymptotic dynamics. The SI is then defined as
\begin{equation}
\mathrm{SI}
=
1-\frac{1}{M}
\sum_{m=1}^{M}
\Theta(\delta-\sigma_m),
\label{eq:eqsi}
\end{equation}
where $\Theta(\cdot)$ is the Heaviside step function and $\delta$ is a prescribed threshold parameter chosen as in Ref.~\cite{gopal2014observation}. By construction, $\mathrm{SI}\approx0$ indicates a spatially coherent state, $\mathrm{SI}\approx1$ corresponds to a fully incoherent state, whereas intermediate values, $0<\mathrm{SI}<1$, reveal the coexistence of coherent and incoherent spatial domains, which is the hallmark of chimera states.

\begin{figure*}[htp!]
\centering
\begin{tikzpicture}
\node[anchor=south west, inner sep=0] (tab) at (0,0) {
 \includegraphics[width=0.9\textwidth]{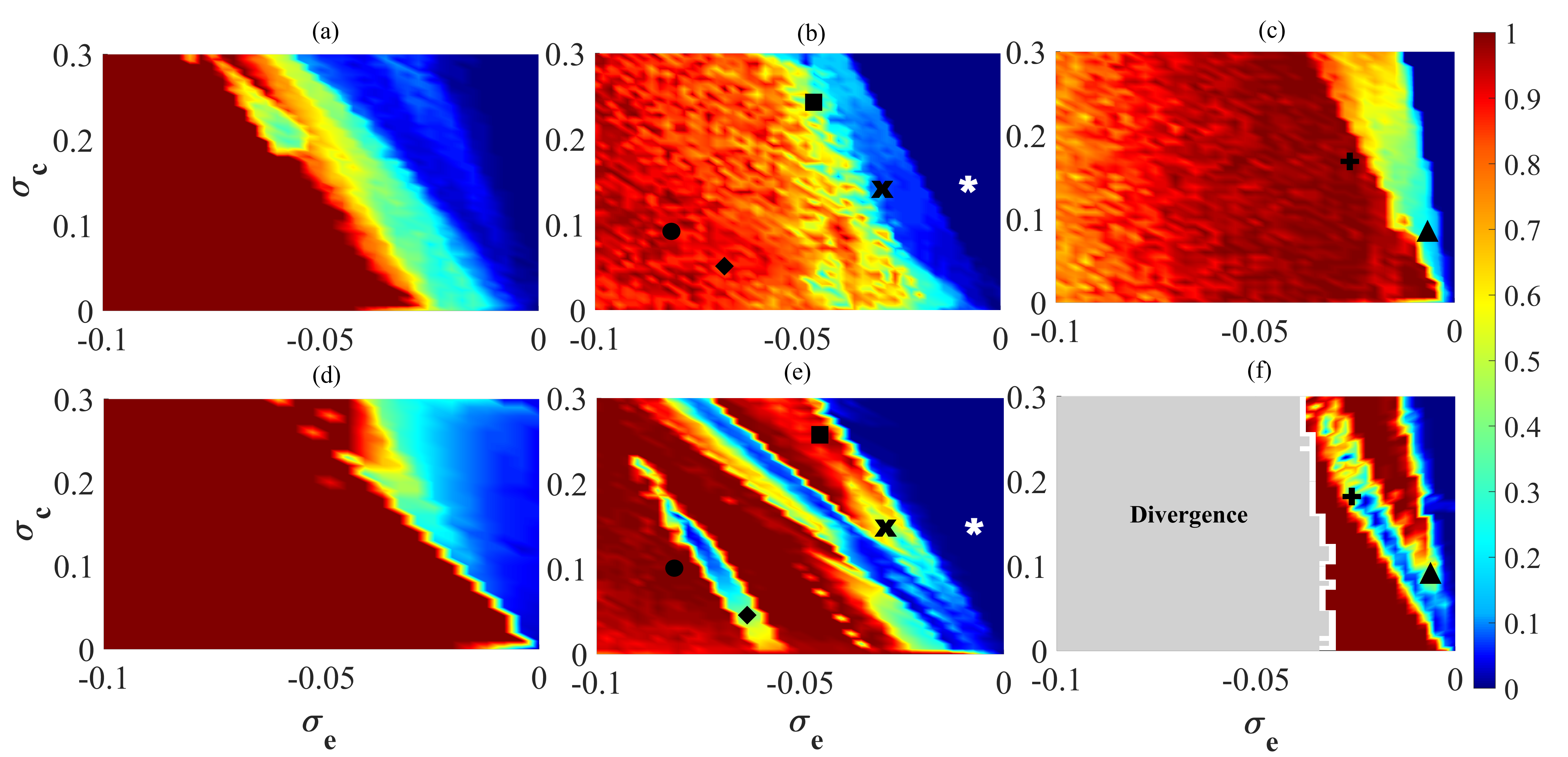}};
\vspace{-5 cm}
\begin{scope}[x={(tab.south east)}, y={(tab.north west)}]
\draw[-{Stealth[length=6pt,width=6pt]}, line width=1.5pt](0.085,0.0) -- (0.90,0.0);
\draw[-{Stealth[length=6pt,width=6pt]}, line width=1.5pt](-0.02,0.20) -- (-0.02,0.90);
\node[anchor=north] at (0.17,-0.03)  {\textit{d=}2};
\node[anchor=north] at (0.5,-0.03)  {\textit{d=}3};
\node[anchor=north] at (0.85,-0.03)  {\textit{d=}4};
\node[anchor=east, rotate=90] at (-0.0325,0.820) {\textit{Hypergraphs}};
\node[anchor=east, rotate=90] at (-0.0325,0.40) {\textit{Projections}};
\end{scope}
\end{tikzpicture}
\caption{ \textbf{Hypergraphs vs. pairwise interactions.} Strength of Incoherence (SI) as a function of the repulsive electrical coupling strength ($\sigma_e$) and attractive chemical coupling strength ($\sigma_c$) for (a--c) genuine higher-order interactions with pure hypergraphs and (d--f) the corresponding pairwise interactions obtained through clique projection of the hypergraphs. From left to right, the hyperedge size is increased from (a,d) $d= 2$, (b,e) $d= 3$, to (c,f) $d= 4$, corresponding to hyperedges containing $d+1$ nodes. The color scale represents the SI value, with SI $\approx 0$ indicating coherent states, SI $\approx 1$ corresponding to fully incoherent dynamics, and intermediate values $0 <$ SI $< 1$ indicating the coexistence of coherent and incoherent domains, characteristic of chimera states. The black symbols superimposed on the phase diagrams highlight selected parameter combinations $(\sigma_e,\sigma_c)$ used to compare the detailed spatiotemporal dynamics of the two network representations [see Figs.~\ref{fig:timecomp}--\ref{fig:5}]. To avoid overloading the caption, the corresponding parameter
values are specified in the captions of the subsequent figures where they are discussed.
}
\label{fig:SI_map1}
\end{figure*}

We use this framework to systematically investigate the collective dynamics of the system described above. Specifically, we examine the combined effects of attractive ($\sigma_c>0$) and repulsive ($\sigma_e<0$) couplings, as well as the role of hyperedge size in shaping the collective dynamics under higher-order interactions (HOIs). Furthermore, this framework provides a natural setting for comparing genuine higher-order dynamics with those arising from the corresponding pairwise networks obtained via clique projection. Fig.~\ref{fig:SI_map1} shows the strength of incoherence phase diagrams in the $(\sigma_e,\sigma_c)$ parameter space for hyperedge sizes $d=2$, $3$, and $4$, where each hyperedge contains $d+1$ nodes (i.e., $3$, $4$, and $5$ nodes, respectively). The upper row [Fig.~\ref{fig:SI_map1}(a--c)] corresponds to the genuine hypergraph framework, while the lower row [Fig.~\ref{fig:SI_map1}(d--f)] presents the results for the associated clique-projected pairwise networks by preserving the same degree nonlinearity used in the hypergraphs coupling terms. Rather than merely comparing higher-order interactions with pairwise interactions, this figure also highlights the combined role of repulsive electrical, and attractive chemical coupling, in the emergence of collective behaviors in neural networks.
For sufficiently strong repulsive coupling, i.e., for large negative values of the electrical coupling strength, the repulsive interaction dominates the attractive coupling and drives the entire network toward a fully incoherent or desynchronized state, characterized by $\mathrm{SI}\to1$ (red region). This behavior is consistently observed both in the presence of HOIs [Fig.~\ref{fig:SI_map1}(a--c)] and for the corresponding projected graphs [Fig.~\ref{fig:SI_map1}(d--f)]. Conversely, increasing the strength of the repulsive electrical coupling progressively disrupts global incoherence, driving the system toward coherent states with $\mathrm{SI}\to 0$. Between these two extremes, an intermediate regime emerges in which the attractive and repulsive interactions partially balance each other, giving rise to the coexistence of coherent and incoherent subpopulations, the hallmark of chimera states. This regime is characterized by intermediate values of the strength of incoherence, $0 < \mathrm{SI} < 1$.

This analysis reveals that increasing the hyperedge order from $d=2$ to $d=4$ substantially reshapes the organization of the dynamical regimes in the parameter space $(\sigma_e,\sigma_c)$. For HOIs, as shown in Fig.~\ref{fig:SI_map1}(a--c), the coherent region progressively contracts as $d$ increases, while incoherent region becomes increasingly dominant, particularly for $d=4$ [see Fig.~\ref{fig:SI_map1}(c)]. According to the literature, this trend is consistent with more general results showing that HOIs can profoundly reorganize synchronization landscapes by generating new transitions, multistability, and modified chimera regimes, rather than simply reproducing the behavior of pairwise networks~\cite{skardal2020higher,ji2024chimera}. It is, however, important to emphasize that the effect of increasing the interaction order is not universal, but depends on the form of the coupling function, the strength of the higher-order coupling, and the underlying network topology~\cite{skardal2020higher}.
The corresponding clique projections [Fig.~\ref{fig:SI_map1}(d--f)] qualitatively reproduce the overall coherence--incoherence competition for observed in the HOIs systems for $d=2$. However, this agreement progressively deteriorates as the interaction order increases, revealing substantial
differences in the organization of the collective dynamical regimes. A particularly striking discrepancy emerges from the comparison between Figs.~\ref{fig:SI_map1}(a) and (d). In the hypergraph representation
[Fig.~\ref{fig:SI_map1}(a)], a broad and well-defined zone characterized by intermediate values of the strength of incoherence ($\mathrm{SI}\approx 0.5$)
extends over the parameter region $\sigma_e<0.2$. This region is associated with chimera states, in which coherent and incoherent subpopulation of neurons coexist. By contrast, this intermediate regime is strongly suppressed and becomes almost inexistent in the corresponding clique-projected network
[Fig.~\ref{fig:SI_map1}(d)]. The present observation is consistent with previous studies such as refs.~\cite{muolo2024phase,kundu2022higher,ghosh2024chimeric} which support that  higher-order interactions can facilitate the emergence and persistence of chimera states, whereas their pairwise representations may considerably reduce the parameter regions supporting such dynamics. The underlying distinction may be justified by the nature of the coupling term: while HOIs account for the simultaneous nonlinear influence of multiple oscillators contribution, clique projections replace these collective contributions with combinations of pairwise interactions, which generally do not reproduce the same dynamical effects~\cite{llabres2026reducibility}. The discrepancies observed in Fig.~\ref{fig:SI_map1}(a) and (d) become even more pronounced when interaction order is increasing. For $d=3$, [see Fig.~\ref{fig:SI_map1}(b,e)] the clique-projected system displays alternating
bands of coherent ($\mathrm{SI}\to 0$), incoherent ($\mathrm{SI}\to 1$), and intermediate dynamical regimes ($0<\mathrm{SI}<1$) that differ markedly from those of the corresponding hypergraph. In the last column [Fig.~\ref{fig:SI_map1}(c,f)] which correspond to $d=4$, a substantial portion of the parameter space exhibits numerically divergent trajectories, indicating a further qualitative departure
between the two descriptions.

\begin{figure*}[htp!]
    \centering
\resizebox{\textwidth}{!}{%
\begin{tikzpicture}
    \def\imgW{8cm}
    \def\imgH{4 cm}
    \def\boxGap{0.15cm}
    \tikzset{
        panelbox/.style={
            draw=black!70,
            line width=1.4pt,
            rounded corners=8pt,
            inner sep=0.20cm,
            anchor=north west
        }
    }
   \node[panelbox] (boxA) at (0,0) {%
        \begin{minipage}[c][\imgH]{\imgW}
            \centering
            \includegraphics[width=1\textwidth]
            {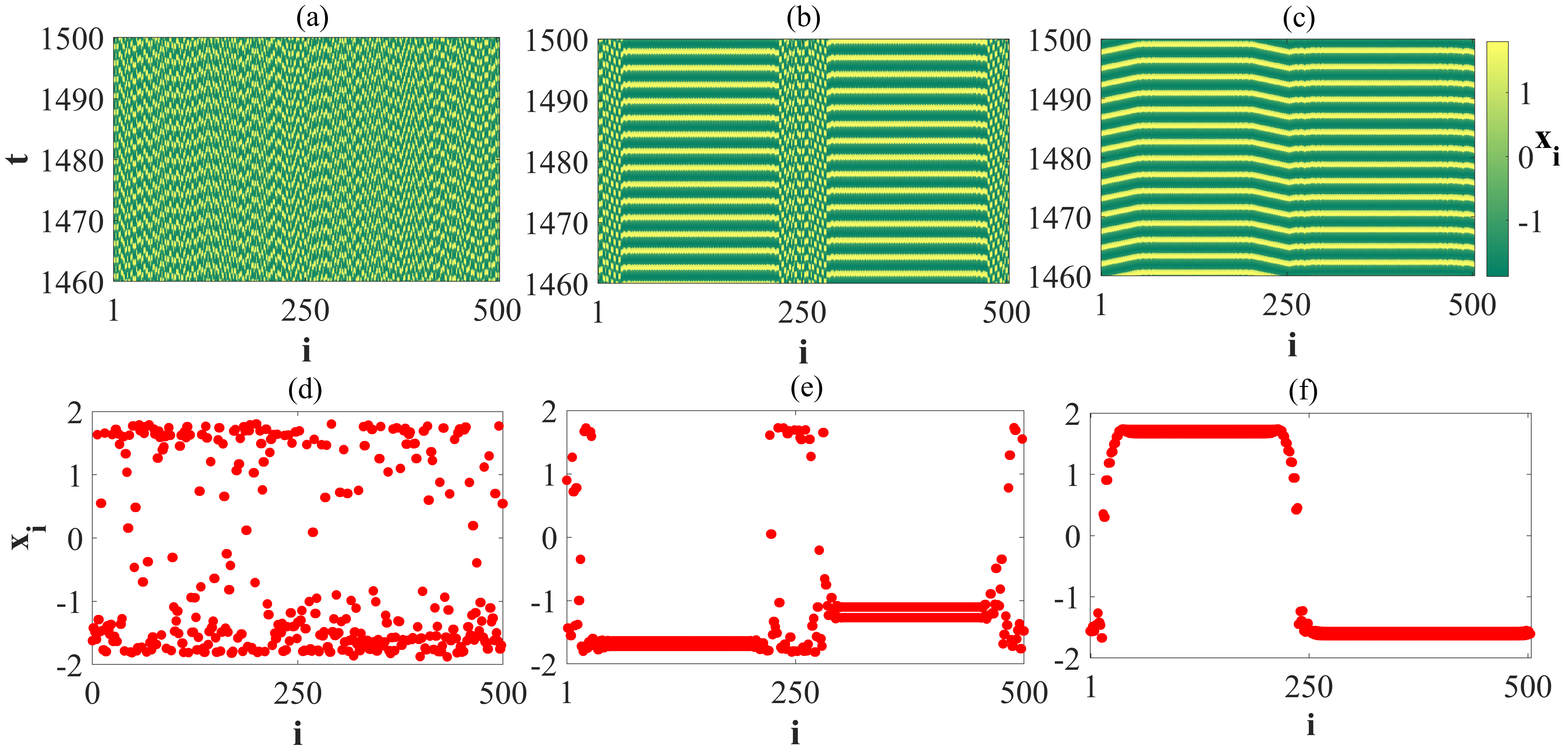}
        \end{minipage}%
    };
    \node[panelbox] (boxB)
        at ([xshift=\boxGap]boxA.north east) {%
        \begin{minipage}[c][\imgH]{\imgW}
            \centering
            \includegraphics[width=\linewidth]
            {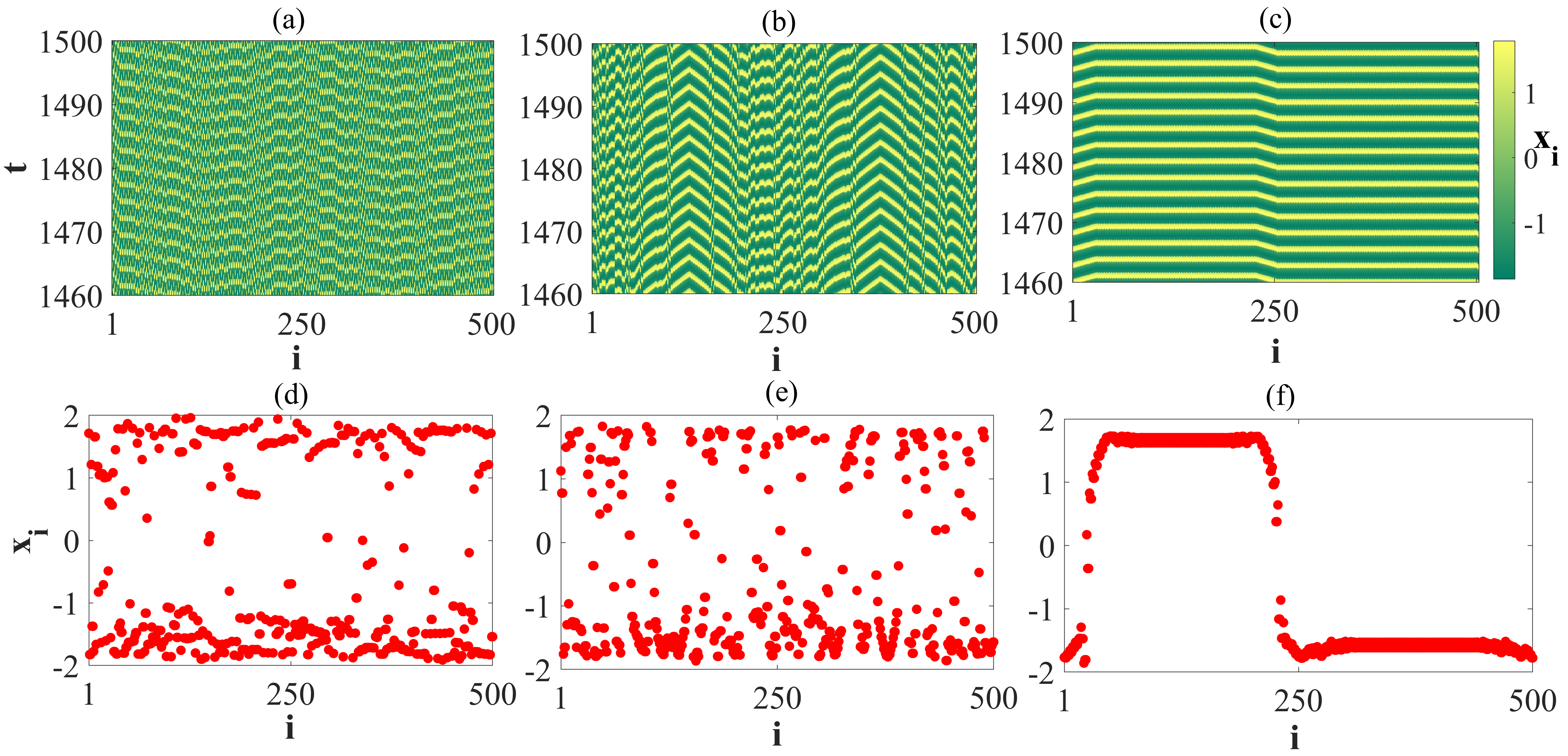}
        \end{minipage}%
    };
    \node[
        anchor=south,
        font=\normalfont\bfseries,
        yshift=0.12cm
    ] at (boxA.north)
    {A. Higher-order interactions};
    \node[
        anchor=south,
        font=\normalfont\bfseries,
        yshift=0.12cm
    ] at (boxB.north)
    {B. Pairwise interactions};
\end{tikzpicture}%
}
    \caption{\textbf{Spatiotemporal Dynamics: Higher-Order vs. Pairwise Interactions.} Comparison of the spatiotemporal dynamics
        obtained for (A) higher-order interactions
        on hypergraphs involving four-body interactions (i.e., $d=3$) and (B) its pairwise interactions on the corresponding clique-projected networks. 
        In both cases, the first row [(a--c)] shows the time series, while the second row [(d--f)] presents the corresponding snapshots taken at $t=1480$.
        In both Panels, from left to right, the columns correspond to three representative parameter combinations $(\sigma_e,\sigma_c)$: (a,d) $(-0.08,0.10)$, marked by a circle ($\bullet$) in Fig.~\ref{fig:SI_map1}, showing incoherent dynamics in both cases; (b,e) $(-0.05,0.28)$, marked by a filled square ($\blacksquare$), exhibiting chimera states in the hypergraph and nearly incoherent dynamics in the projected graph; and (c,f) $(-0.01,0.15)$, marked by a bold asterisk ($\bm{\ast}$), corresponding to coherent dynamics in both cases.
    }
    \label{fig:timecomp}
\end{figure*}

To further substantiate the results presented in Fig.~\ref{fig:SI_map1}, we illustrate in Fig.~\ref{fig:timecomp} the spatiotemporal evolution
of the oscillators and their corresponding instantaneous snapshots plotted at $t=1480$. Therefore, we compare the dynamics of the original hypergraph with
four-body interactions ($d=3$) and those of its corresponding clique-projected network using pairwise representation for three representative combinations of the coupling parameters $(\sigma_e,\sigma_c)$, indicated in Fig.~\ref{fig:SI_map1}. For $(\sigma_e,\sigma_c)=(-0.08,0.10)$, [see Figs.~\ref{fig:timecomp}A(a,d) and B(a,d)] both systems, i.e., the original hypergraph and its corresponding clique-projected networks exhibit incoherent patterns. For $(\sigma_e,\sigma_c)=(-0.05,0.28)$ marked by a filled square in Fig.~\ref{fig:SI_map1}, a pronounced difference is observed between Figs.~\ref{fig:timecomp} A(b,e) and B(b,e). In the higher-order system [Fig.~\ref{fig:timecomp} A(b,e)], the spatiotemporal evolution and the corresponding snapshot reveal the coexistence of coherent and incoherent subpopulations of neurons, a hallmark of chimera states. By contrast, the corresponding clique-projected network [Fig.~\ref{fig:timecomp} B(b,e)] exhibits a markedly different spatiotemporal organization. Although regular wave-like patterns are still observed, the instantaneous spatial distribution
is considerably more dispersed, and the clear separation between coherent and incoherent domains observed in the hypergraph is no longer preserved.
Finally, an interesting case is illustrated in Figs.~\ref{fig:timecomp} A(c,f) and B(c,f), corresponding to $(\sigma_e,\sigma_c)=(-0.01,0.15)$ and marked by a star in Fig.~\ref{fig:SI_map1}. The results show that both systems exhibit coherent spatiotemporal patterns, highlighting a qualitative equivalence between higher-order and pairwise interactions in this particular dynamical regime.

\begin{figure*}[htp!]
    \centering
    \resizebox{\textwidth}{!}{%
    \begin{tikzpicture}
        \def\boxW{8.2cm}
        \def\boxH{6.1cm}
        \def\boxGap{0.35cm}
        \def\imgW{3.7cm}
        \tikzset{
            panelbox/.style={
                draw=black!70,
                line width=1.4pt,
                rounded corners=8pt,
                inner sep=0.20cm,
                anchor=north west
            }
        }
        \node[panelbox] (boxA) at (0,0) {%
            \begin{minipage}[c][\boxH]{\boxW}
                \centering
                \setlength{\tabcolsep}{3pt}
                \renewcommand{\arraystretch}{1.05}
                \begin{tabular}{cc}
                    (a) & (b) \\
                    \includegraphics[width=\imgW]{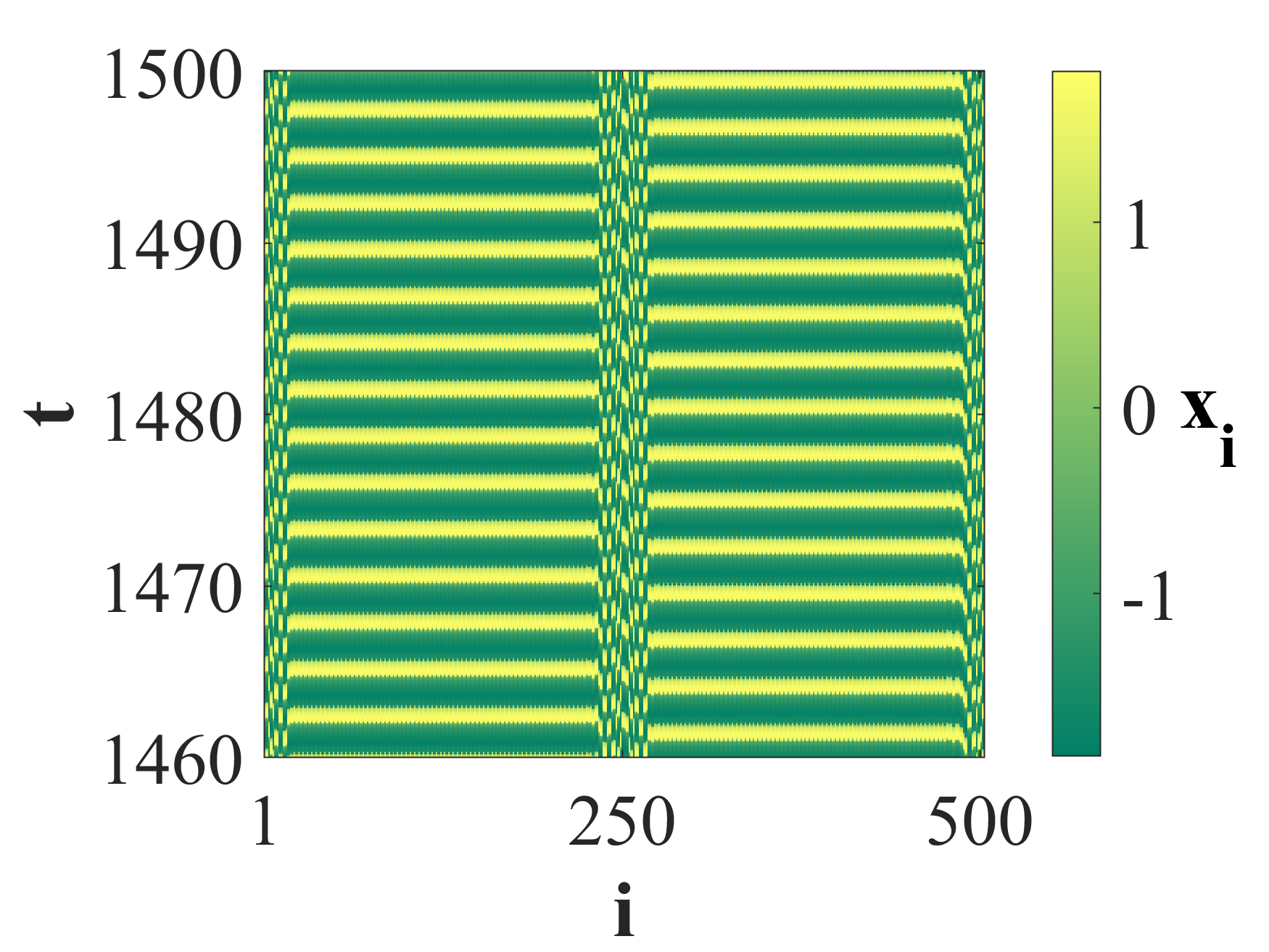} &
                    \includegraphics[width=\imgW]{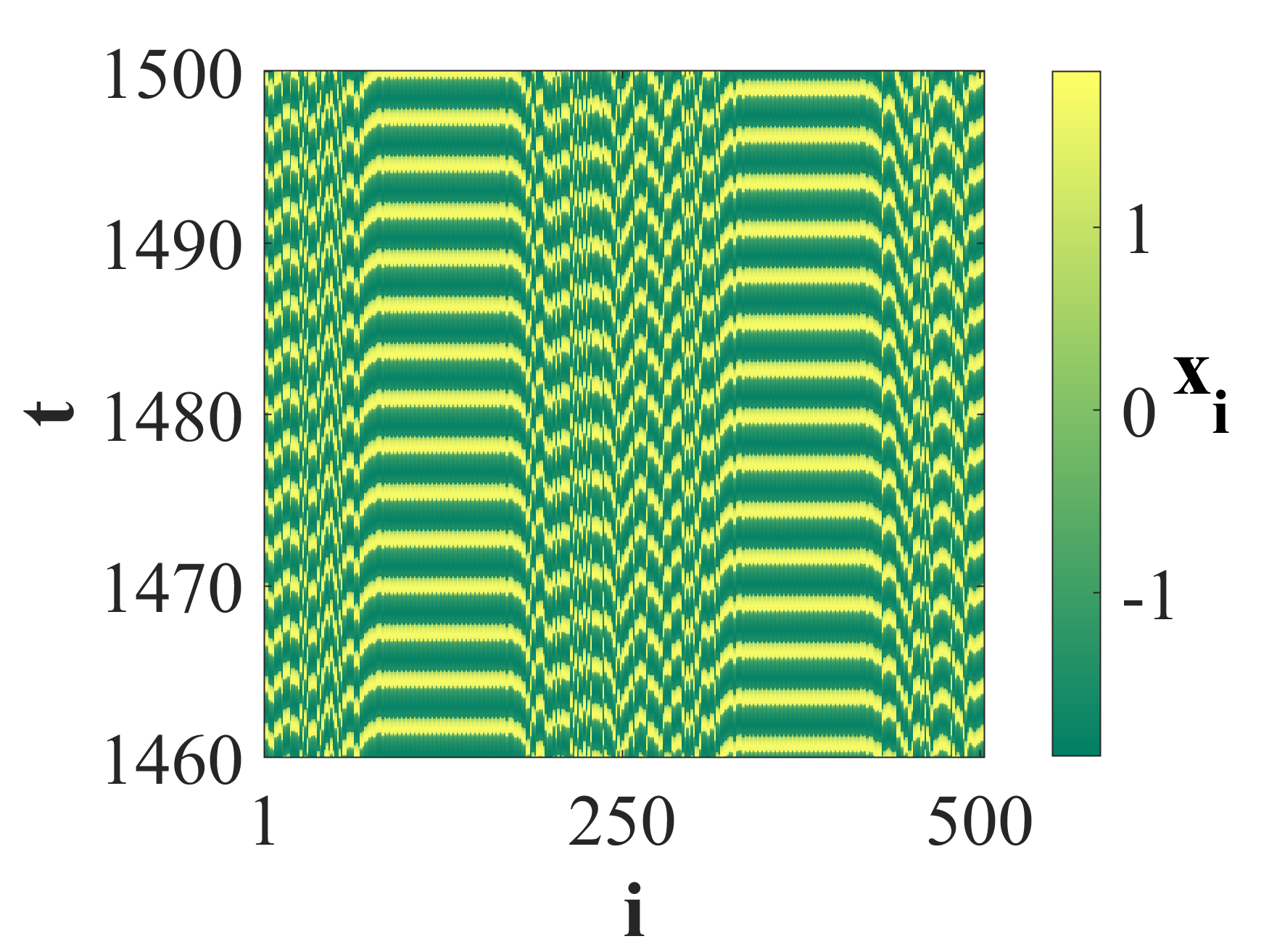} \\
                    (c) & (d) \\
                    \includegraphics[width=\imgW]{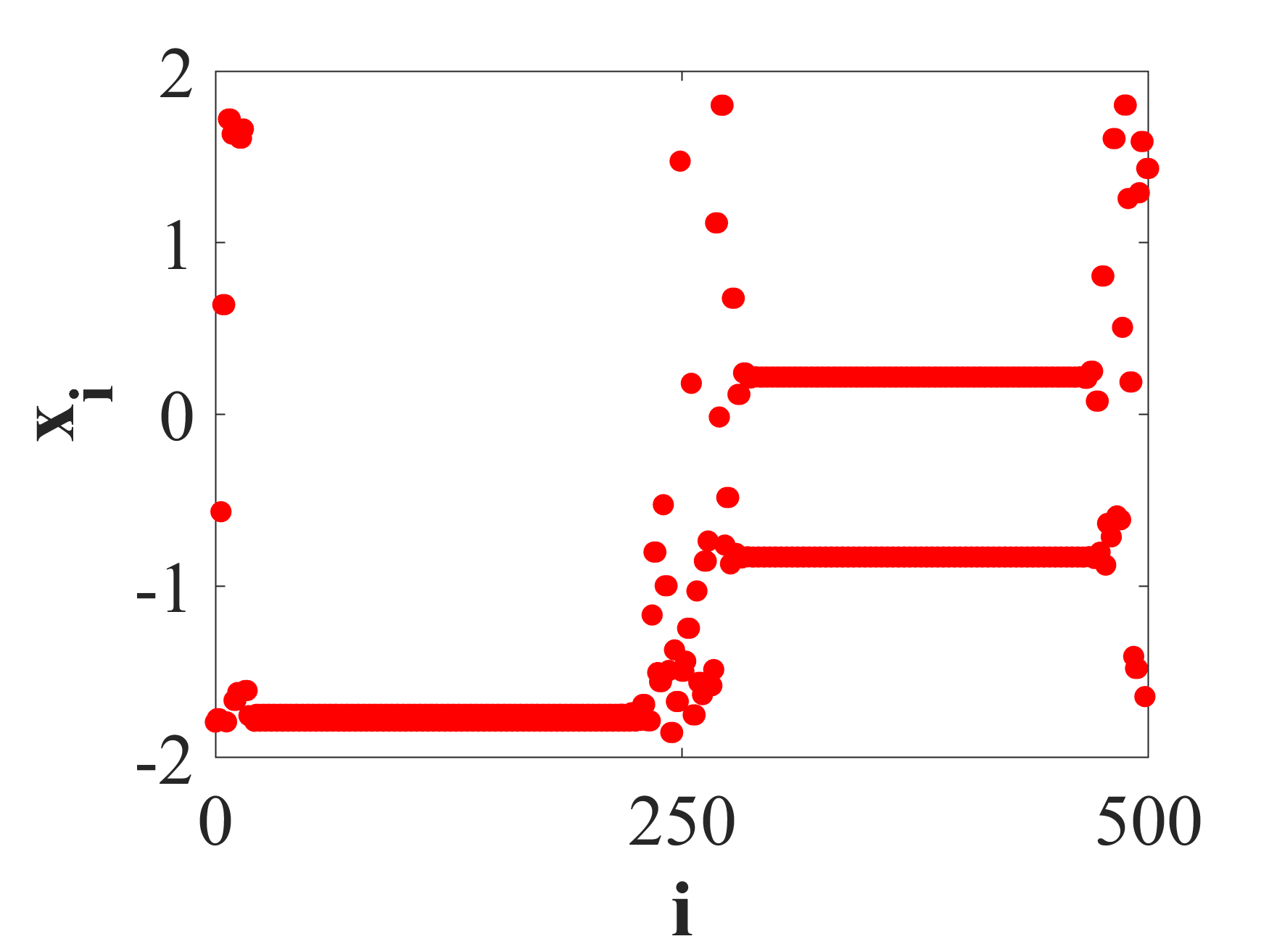} &
                    \includegraphics[width=\imgW]{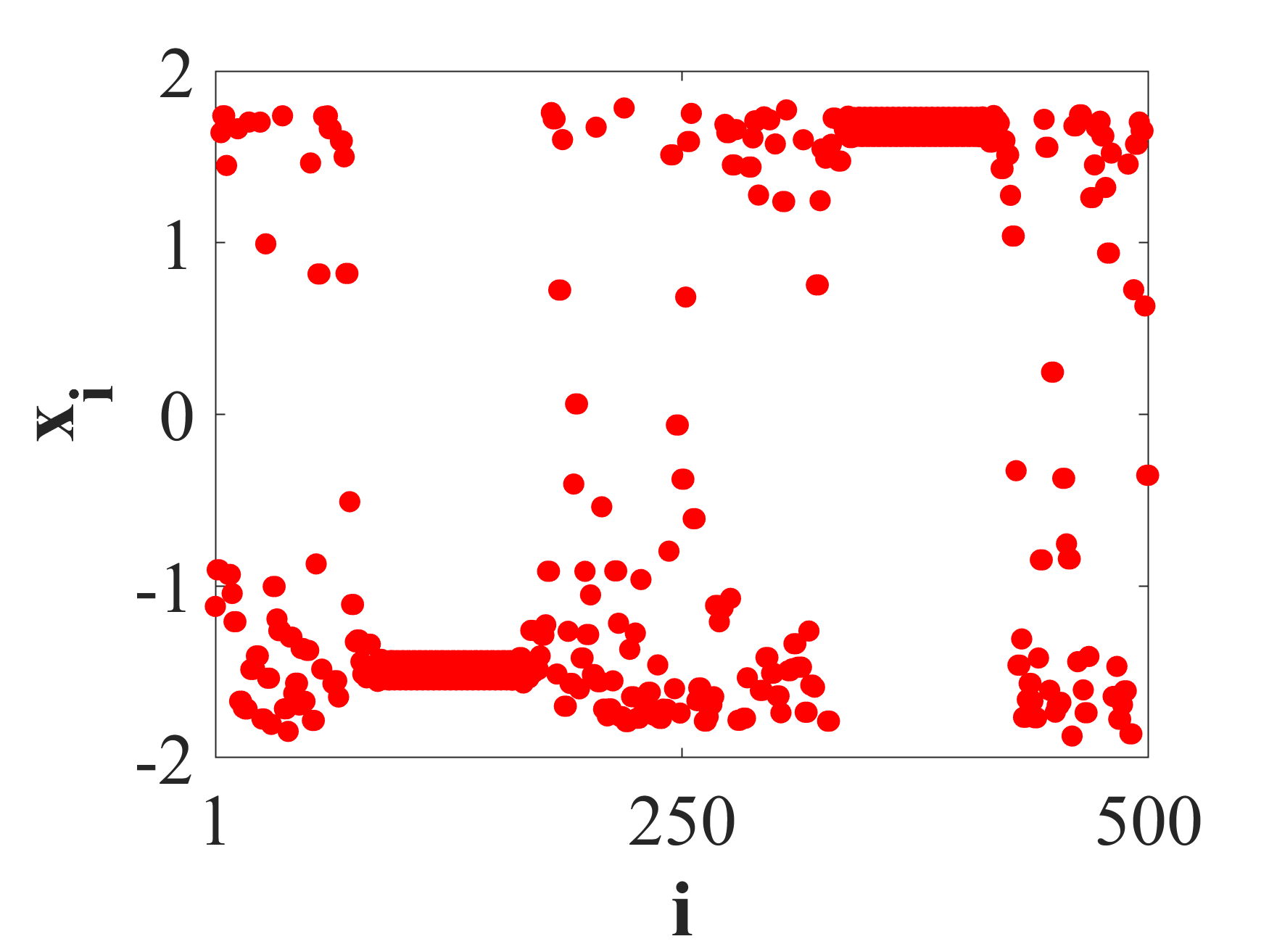}
                \end{tabular}
            \end{minipage}%
        };
        \node[panelbox] (boxB) at ([xshift=\boxGap]boxA.north east) {%
            \begin{minipage}[c][\boxH]{\boxW}
                \centering
                \setlength{\tabcolsep}{3pt}
                \renewcommand{\arraystretch}{1.05}
                \begin{tabular}{cc}
                    (a) & (b) \\
                    \includegraphics[width=\imgW]{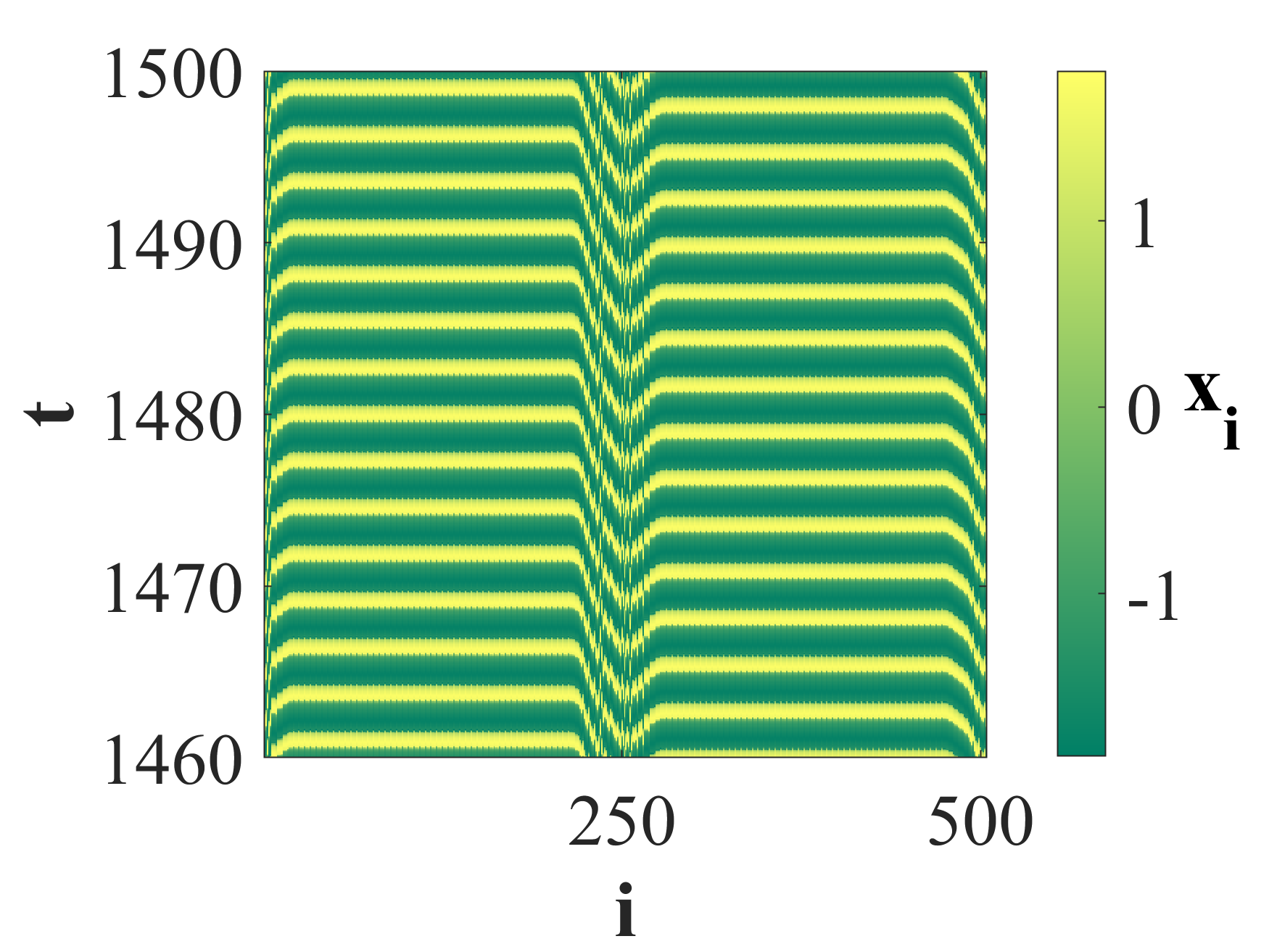} &
                    \includegraphics[width=\imgW]{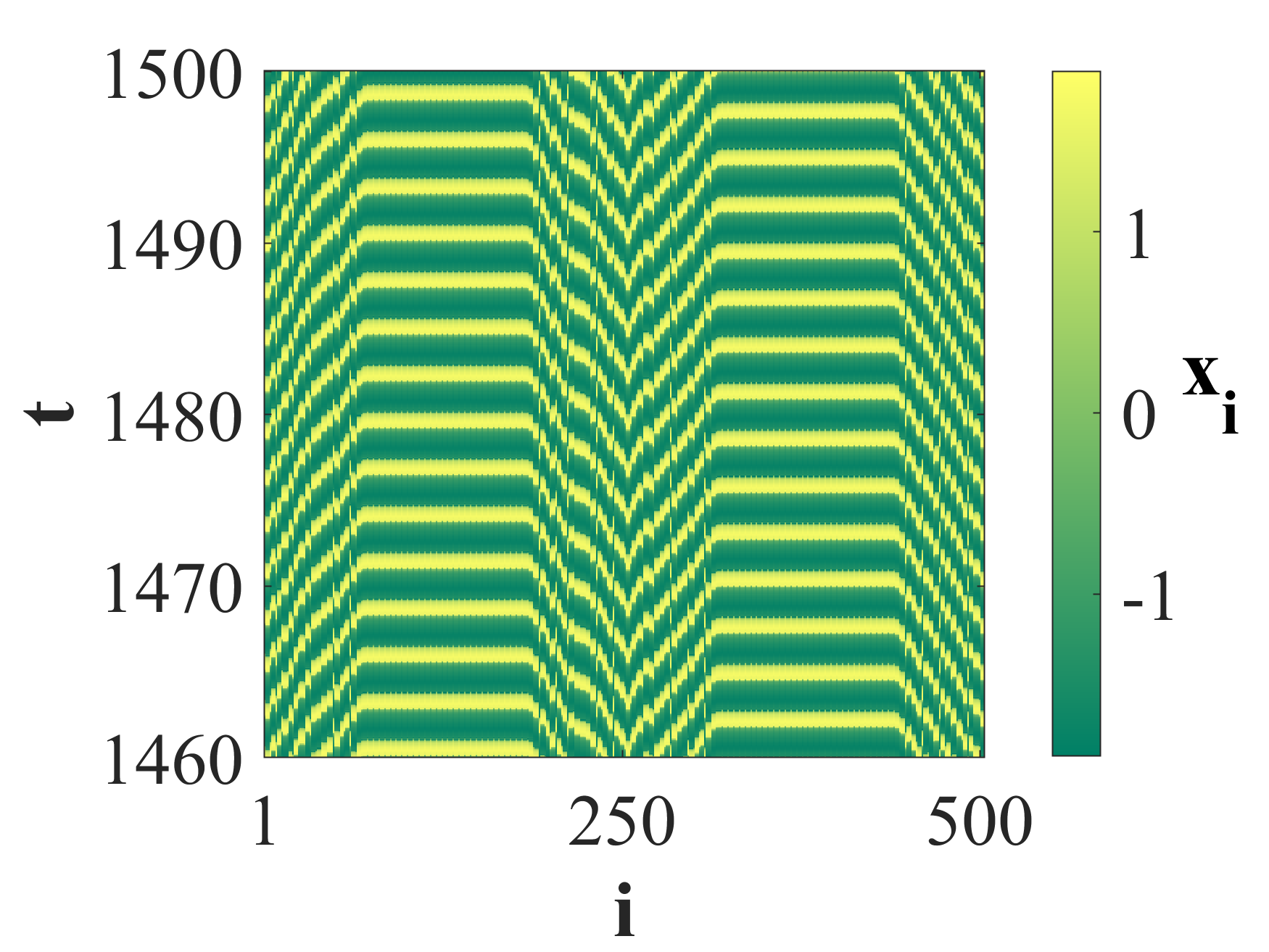} \\
                    (c) & (d) \\
                    \includegraphics[width=\imgW]{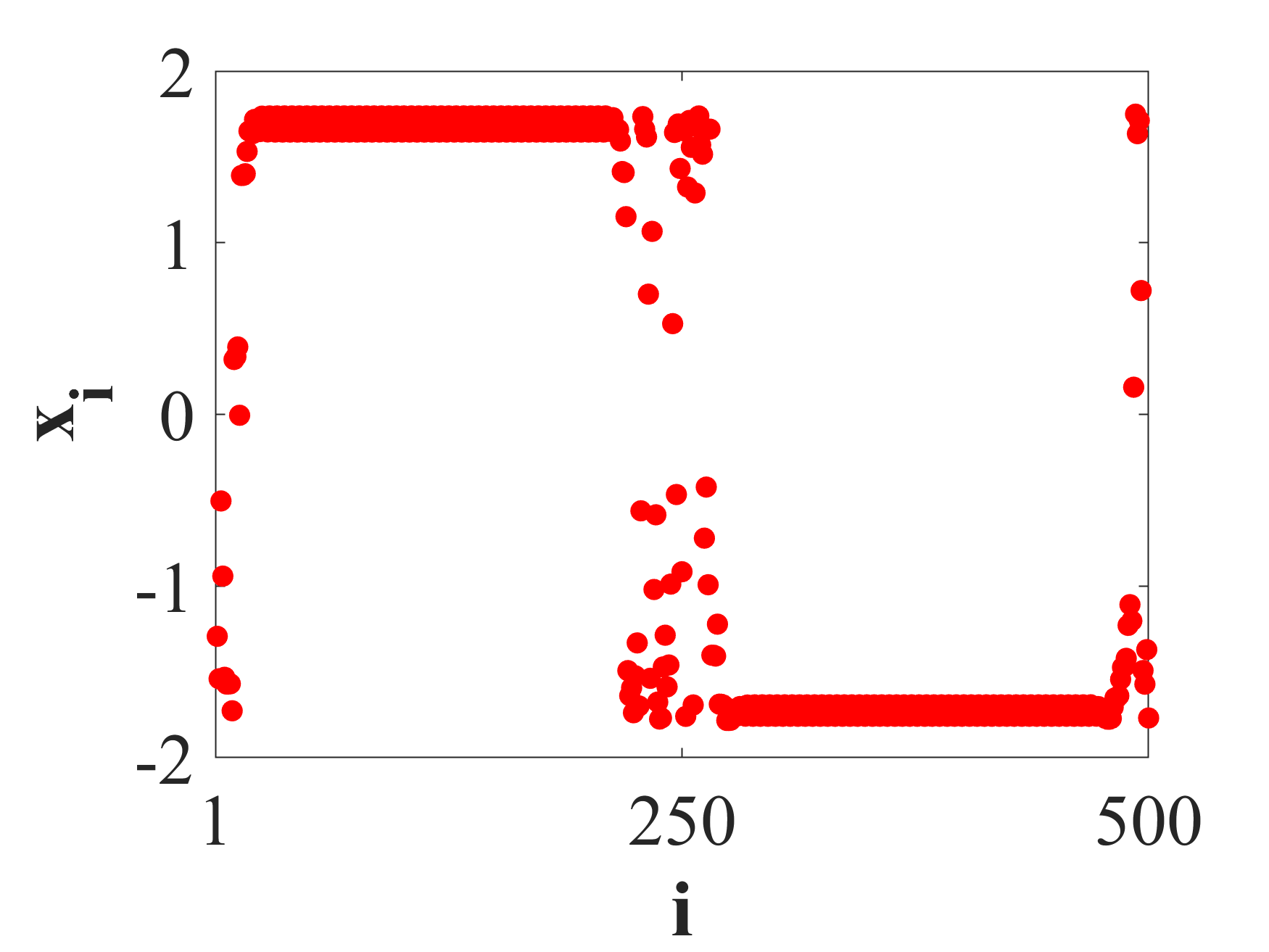} &
                    \includegraphics[width=\imgW]{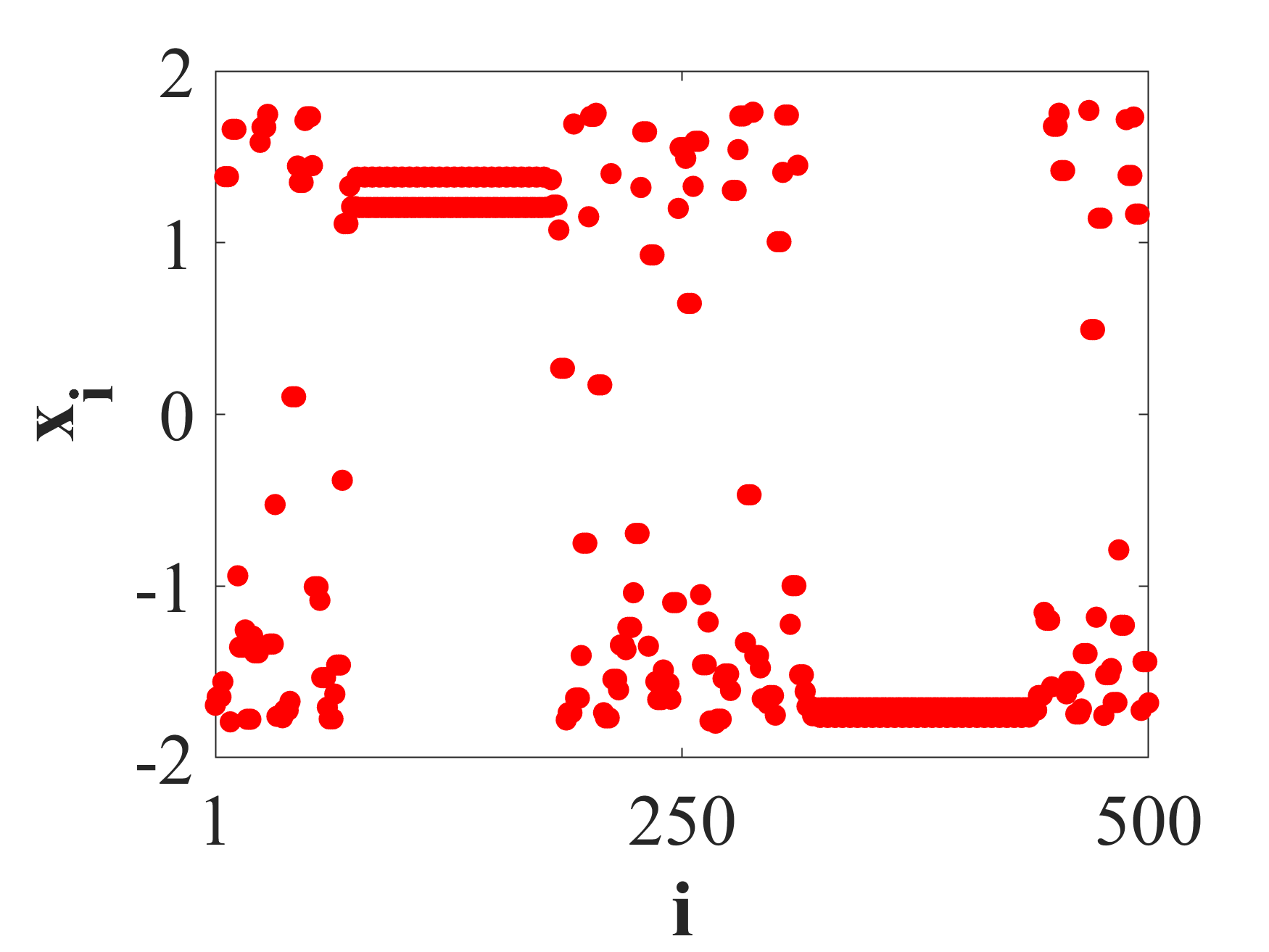}
                \end{tabular}
            \end{minipage}%
        };
        \node[
            anchor=south,
            font=\normalfont\bfseries,
            yshift=0.12cm
        ] at (boxA.north)
        {A: $d=3$};
        \node[
            anchor=south,
            font=\normalfont\bfseries,
            yshift=0.12cm
        ] at (boxB.north)
        {B: $d=4$};
    \end{tikzpicture}%
    }   
    \caption{\textbf{Persistence of chimera states under clique projection.} Comparison of the collective dynamics of hypergraphs and their corresponding clique-projected networks for : (A) three-body $d=3$, and (B) four-body $d=4$ interactions. The coupling parameters are fixed at $(\sigma_e,\sigma_c)=(-0.03,0.15)$ for panel A and $(\sigma_e,\sigma_c)=(-0.005,0.08)$ for panel B, corresponding to the parameter combinations marked by the bold plus symbol ($\bm{\times}$) and the filled triangle ($\blacktriangle$), respectively, in Fig.~\ref{fig:SI_map1}.
 In each panels, (a,c) correspond to the original hypergraph, while panels (b,d) represent its clique-projected counterpart. (a,b) illustrate the spatiotemporal evolution of the activator variable $x_i(t)$, whereas panels (c,d) show the corresponding spatial snapshots captured at $t=1480$. The persistence of coexisting coherent and incoherent domains in both representations demonstrates that chimera states can survive clique projection for both three- and four-body interactions.
    }
    \label{fig:timecomp2}
\end{figure*}

Fig.~\ref{fig:timecomp2} reveals a remarkable dynamical correspondence between the original hypergraphs, with higher-order interactions and their corresponding clique-projected networks, with pairwise interactions for both three-body interactions [Fig.~\ref{fig:timecomp2}A.(a--d)] at $(\sigma_e,\sigma_c)=(-0.03,0.15)$ and four-body interactions
[Fig.~\ref{fig:timecomp2}B.(a--d)] at $(\sigma_e,\sigma_c)=(-0.005,0.08)$. Specifically, the first column  of each panel illustrates the spatiotemporal evolution of the activator variable $x_i(t)$ and the snapshots taken at $t=1480$ of the higher-order interactions cases for three- and four-body interactions, respectively, while the second column display the dynamics of their corresponding clique-projected networks. The striking similarities between the spatiotemporal patterns obtained from both representations suggest that, under certain coupling conditions, clique projection can effectively reproduce the collective dynamics generated by genuine higher-order interactions. However, this dynamical correspondence does not hold universally across the explored parameter space. Indeed, the interplay between attractive coupling $\sigma_c>0$ and repulsive coupling $\sigma_e<0$, considered in the present analyses give rise to a particularly rich and nontrivial relationship between higher-order interactions and the emergence of chimera states. Our results indicate that the competition between both coupling strength frequently favors the emergence of chimera states in hypergraph more prominently than in their corresponding clique-projected networks [Fig.~\ref{fig:timecomp}]. This investigations also present an interesting and opposite scenario, in which opposite behavior is also observed. In certain parameter spaces, the combination of both coupling strength $(\sigma_e, \sigma_c)$ favors the projected networks to develop a more pronounced spatially heterogeneous patterns, sometimes exhibiting substantially broader incoherent domains than their higher-order counterparts [Fig.~\ref{fig:timecomp2}].

\begin{figure*}[htp!]
    \centering
    \resizebox{\textwidth}{!}{%
    \begin{tikzpicture}
        \def\boxW{8.2cm}
        \def\boxH{6.1cm}
        \def\boxGap{0.35cm}
        \def\imgW{3.7cm}
        \tikzset{
            panelbox/.style={
                draw=black!70,
                line width=1.4pt,
                rounded corners=8pt,
                inner sep=0.20cm,
                anchor=north west
            }
        }
        \node[panelbox] (boxA) at (0,0) {%
            \begin{minipage}[c][\boxH]{\boxW}
                \centering
                \setlength{\tabcolsep}{3pt}
                \renewcommand{\arraystretch}{1.05}
                \begin{tabular}{cc}
                    (a) & (b) \\
                    \includegraphics[width=\imgW]{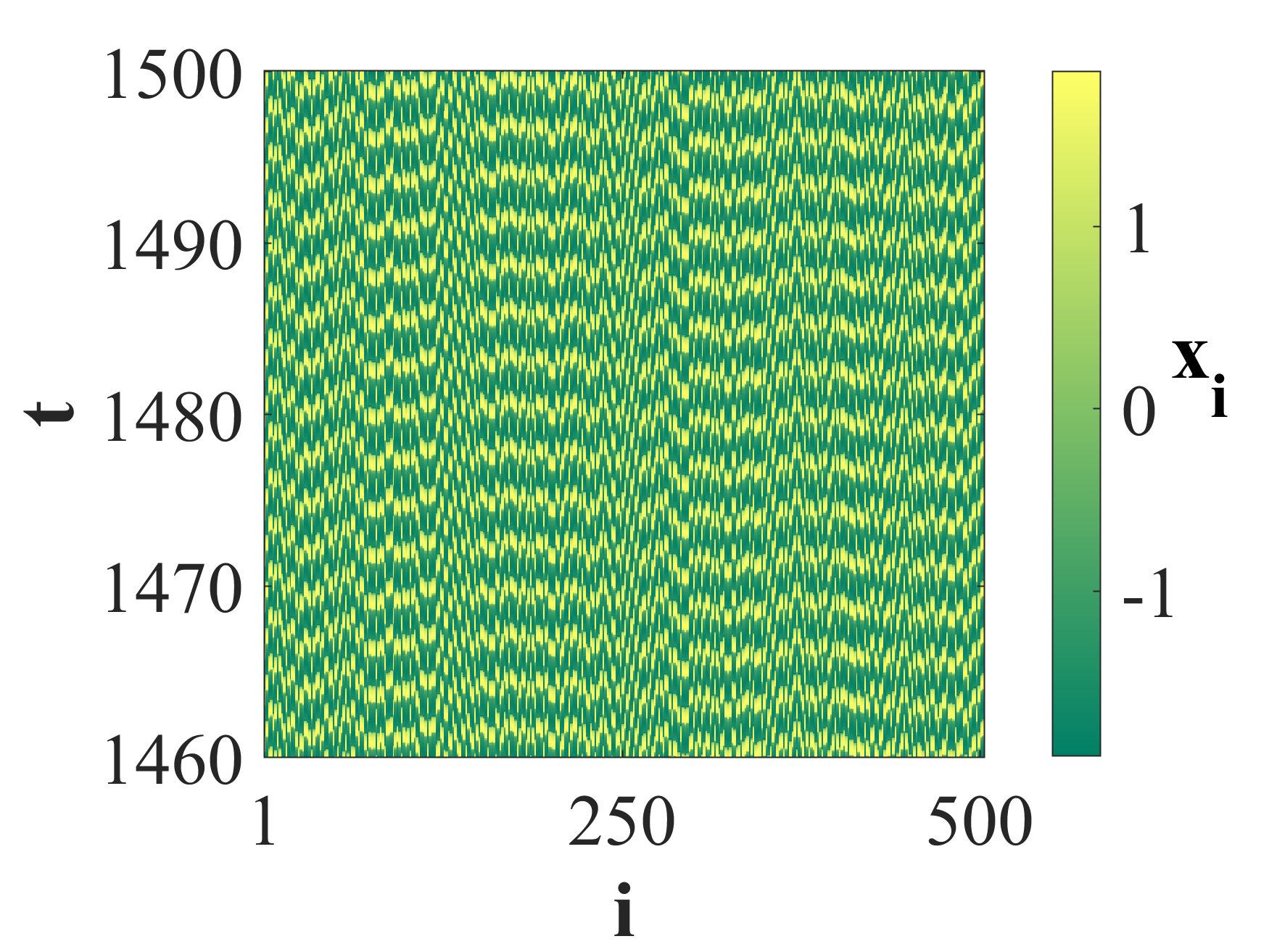} &
                    \includegraphics[width=\imgW]{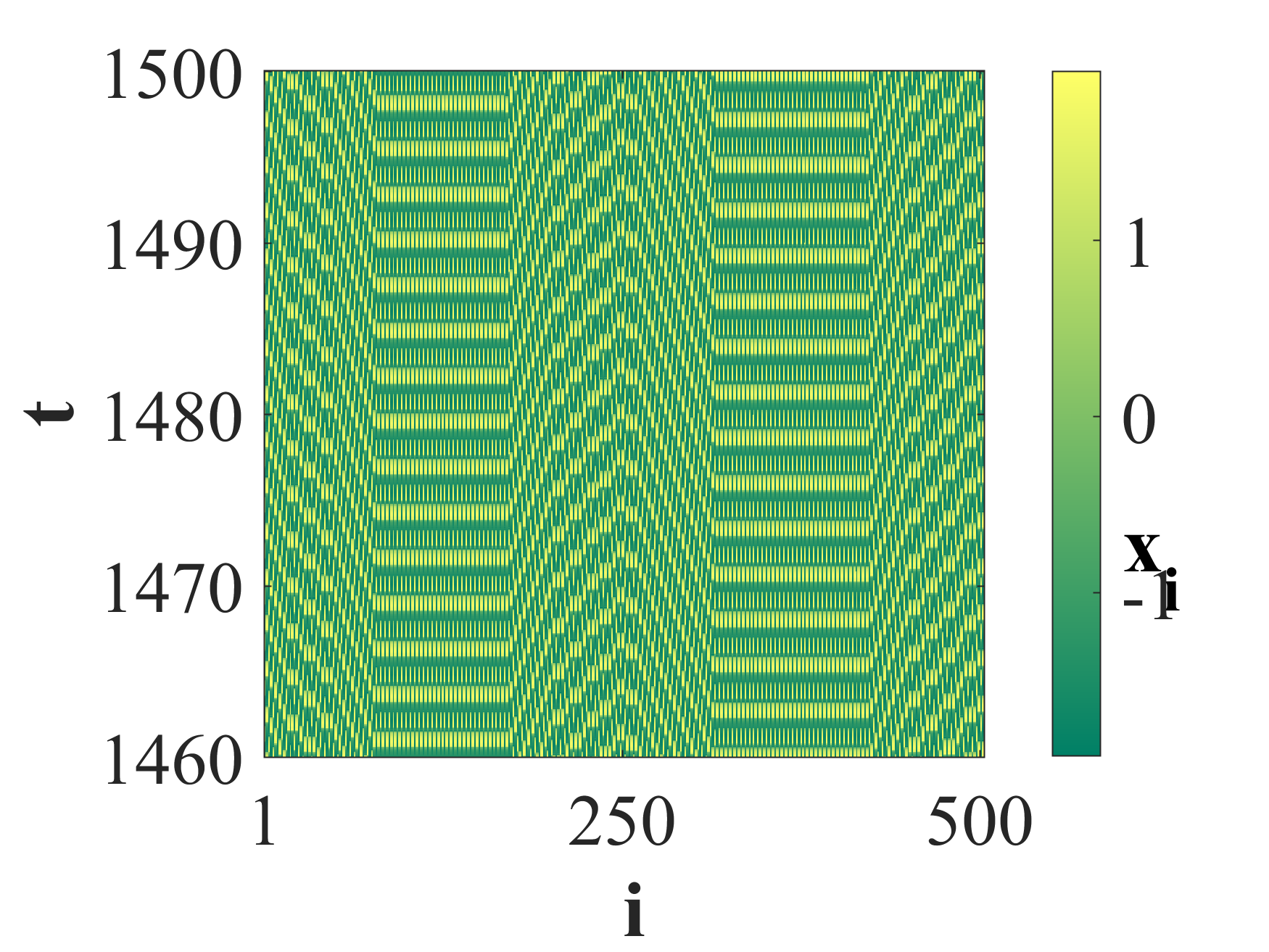} \\
                    (c) & (d) \\
                    \includegraphics[width=\imgW]{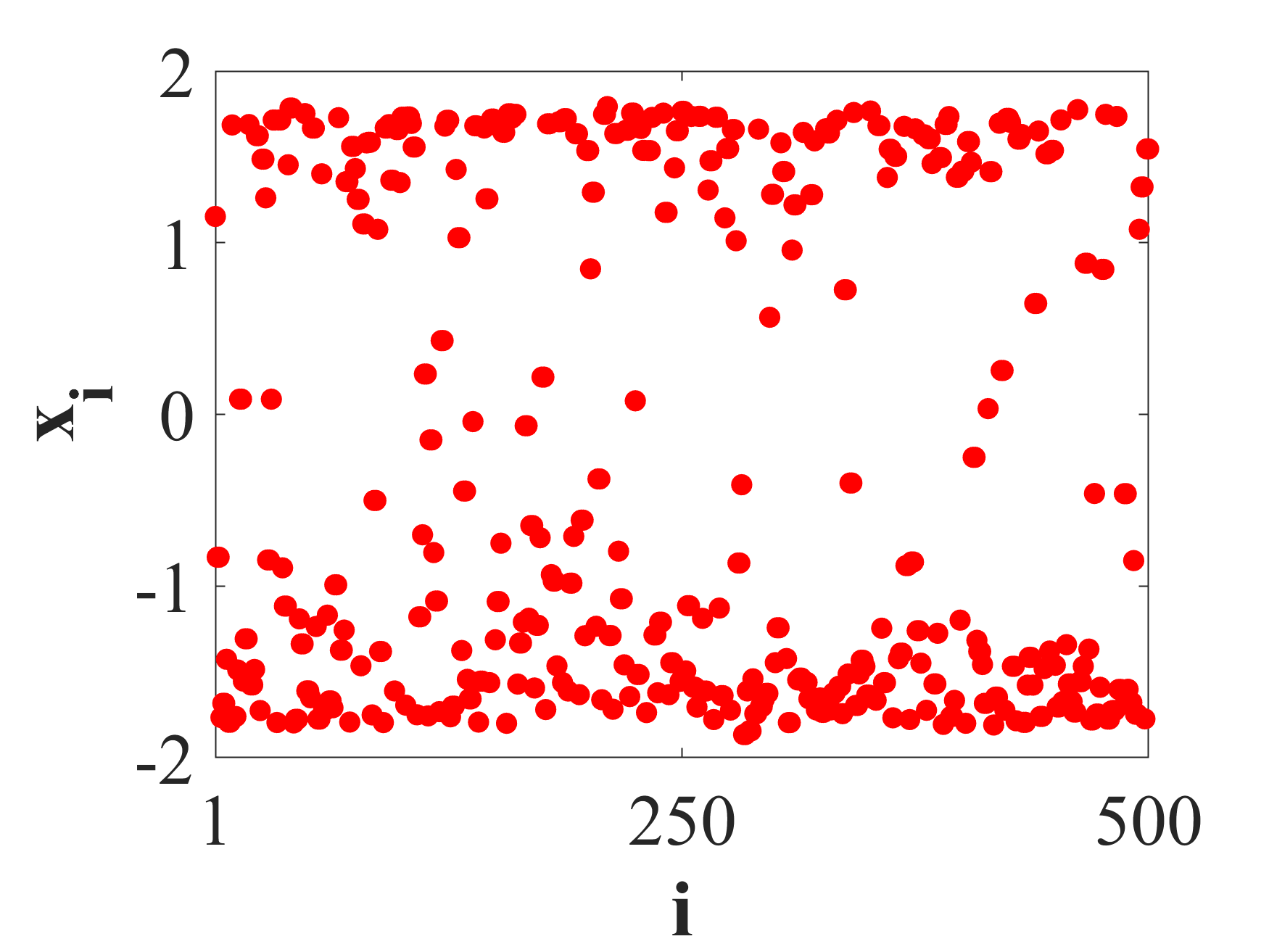} &
                    \includegraphics[width=\imgW]{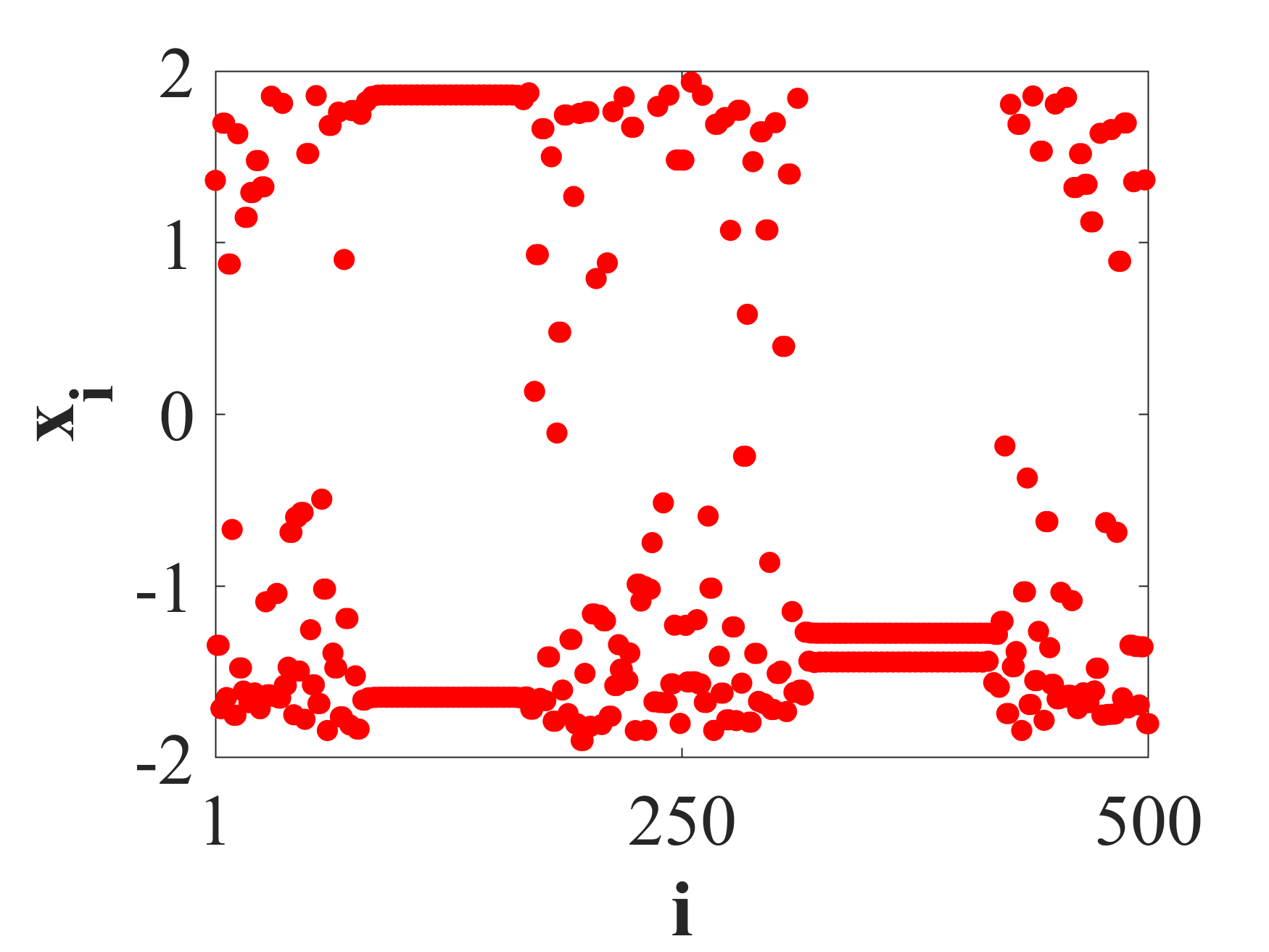}
                \end{tabular}
            \end{minipage}%
        };
        \node[panelbox] (boxB)
            at ([xshift=\boxGap]boxA.north east) {%
            \begin{minipage}[c][\boxH]{\boxW}
                \centering
                \setlength{\tabcolsep}{3pt}
                \renewcommand{\arraystretch}{1.05}
                \begin{tabular}{cc}
                    (a) & (b) \\
                    \includegraphics[width=\imgW]{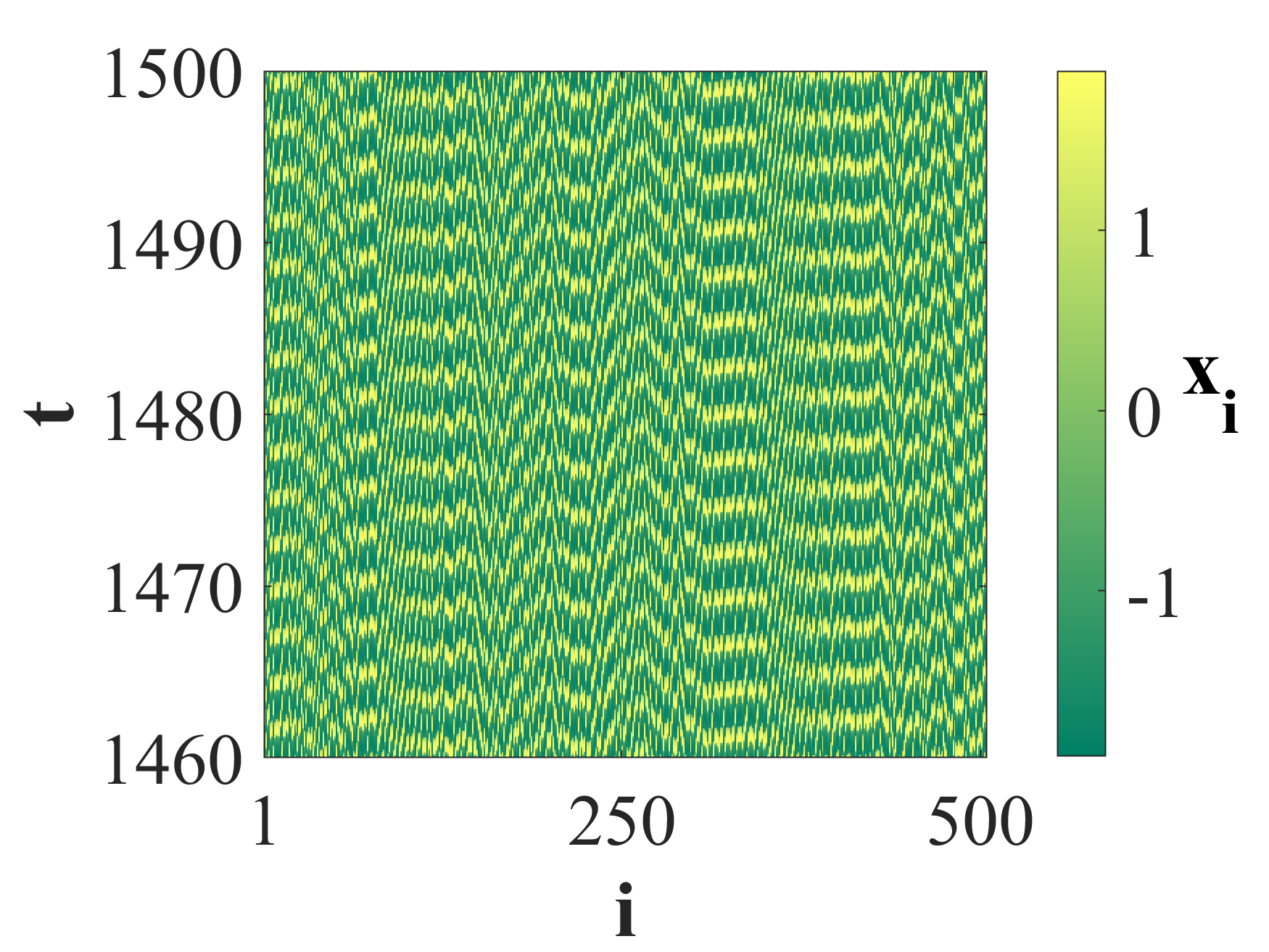} &
                    \includegraphics[width=\imgW]{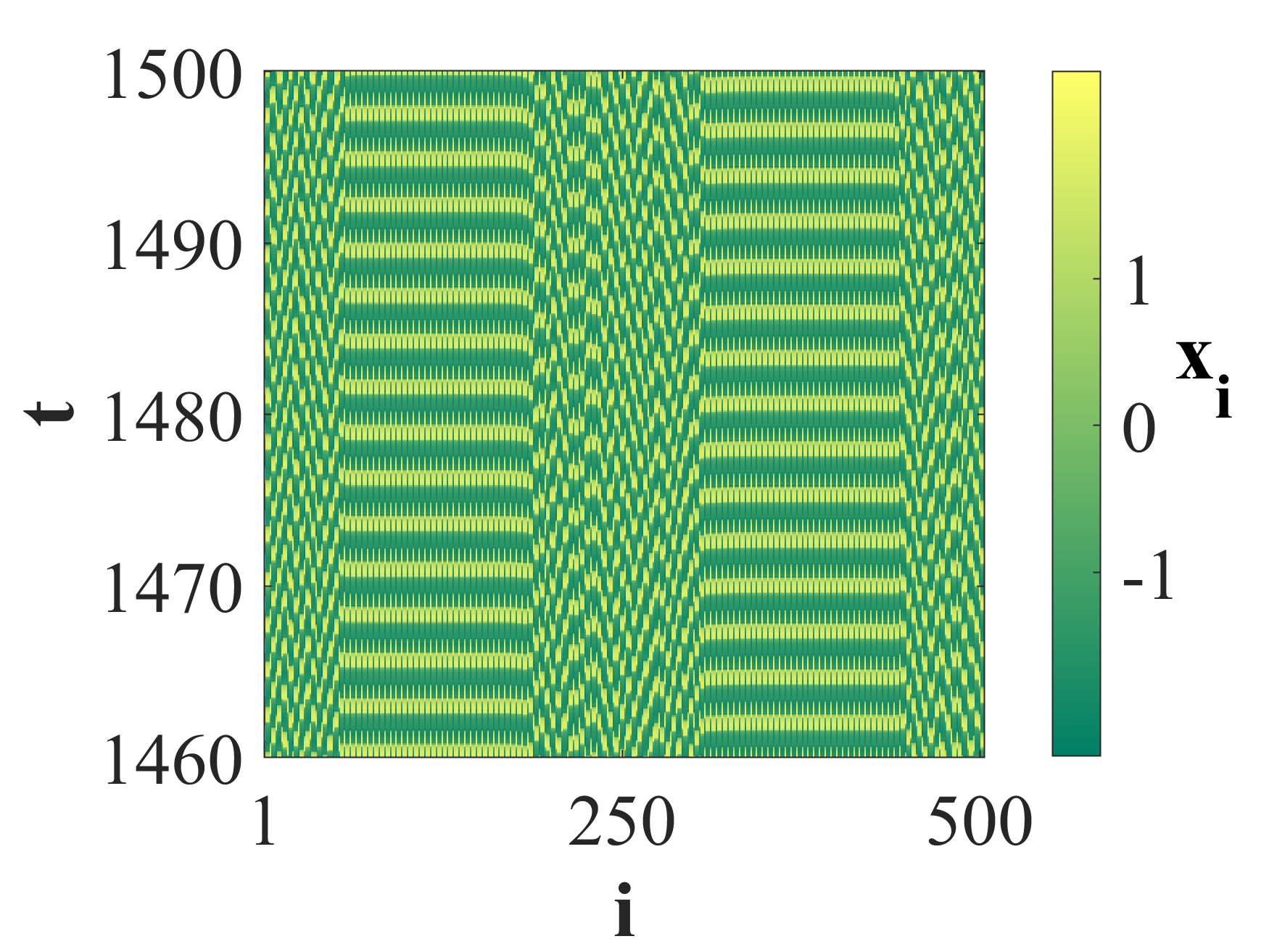} \\
                    (c) & (d) \\
                    \includegraphics[width=\imgW]{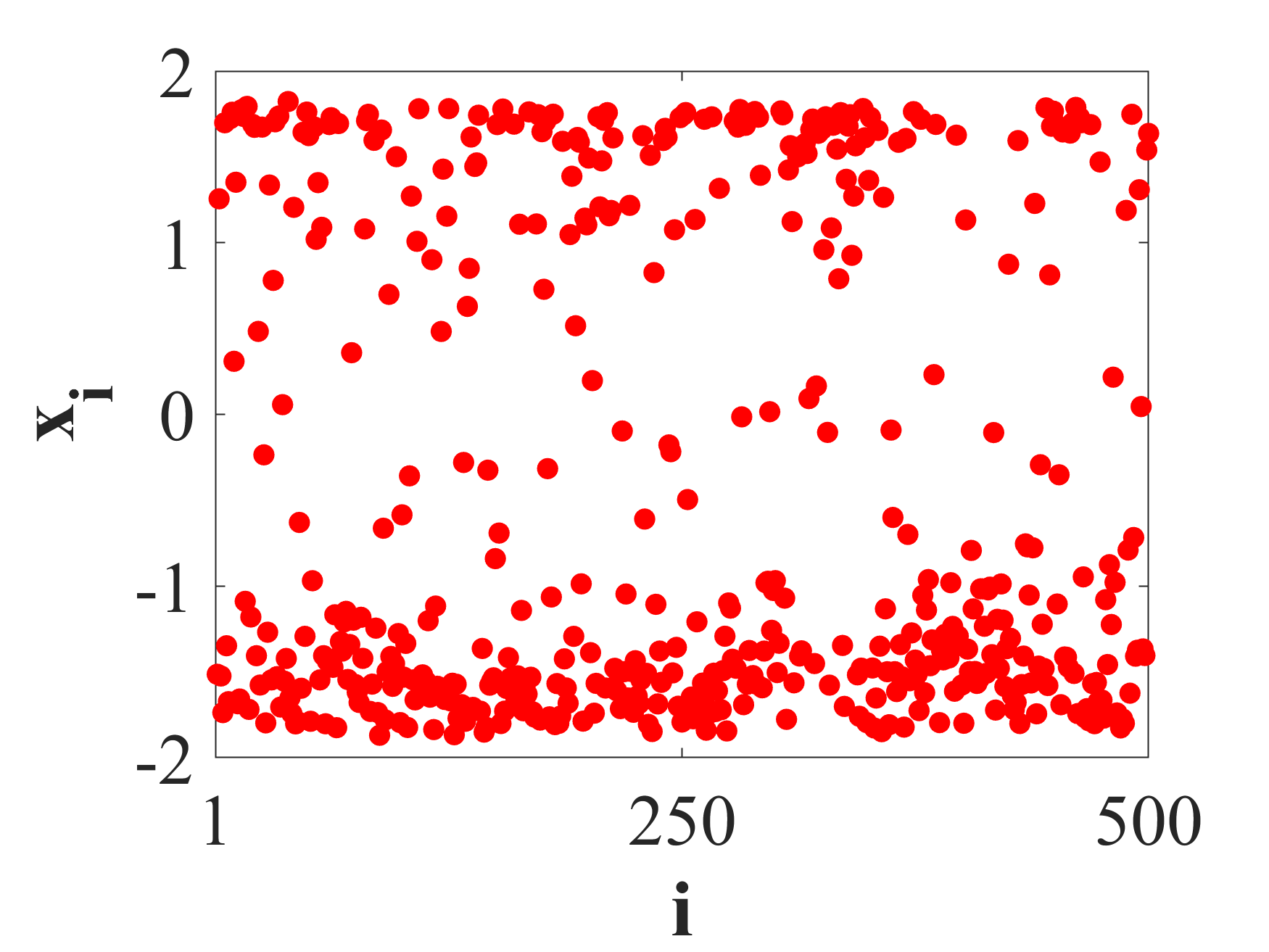} &
                    \includegraphics[width=\imgW]{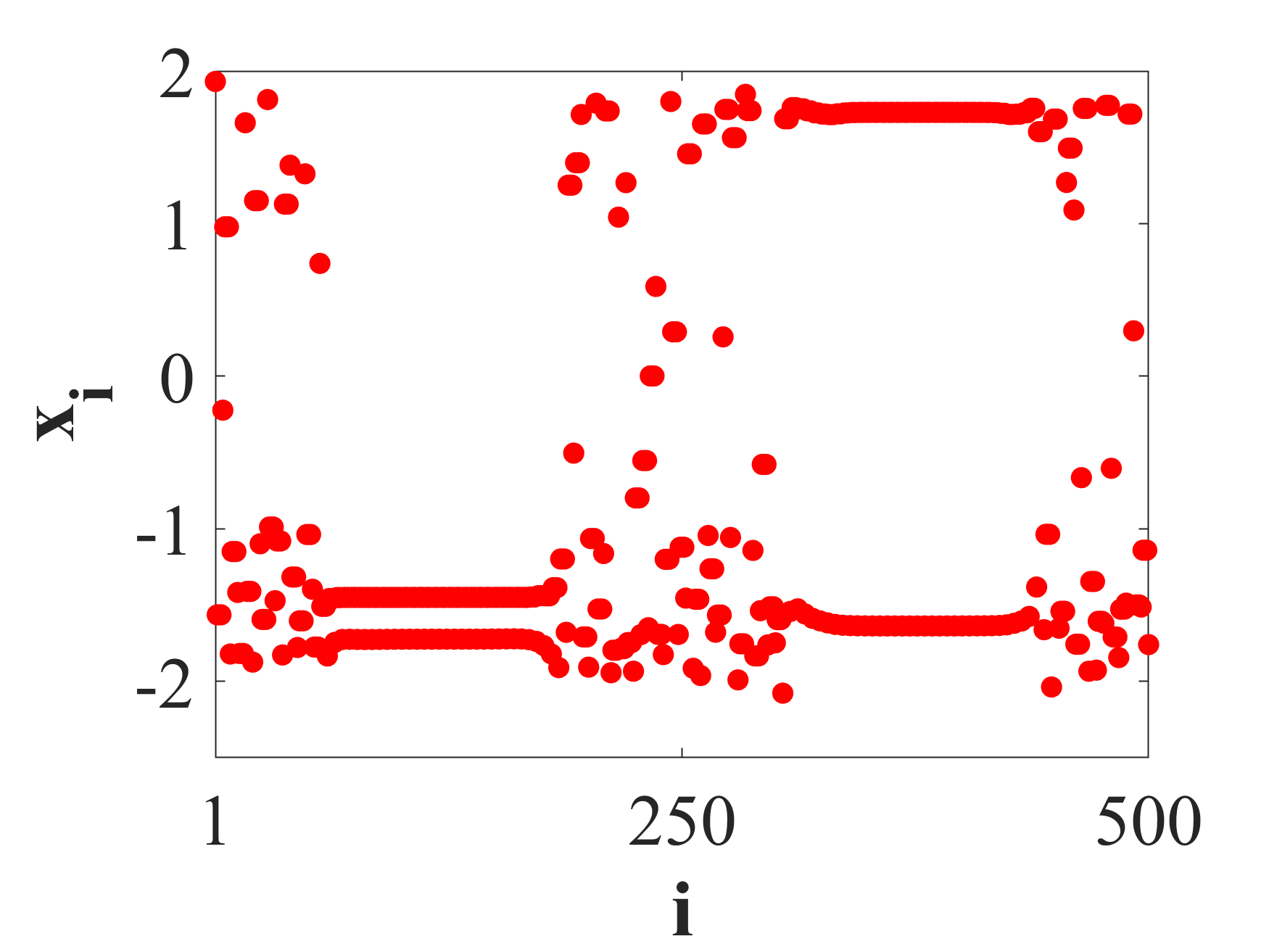}
                \end{tabular}
            \end{minipage}%
        };
        \node[
            anchor=south,
            font=\normalfont\bfseries,
            yshift=0.12cm
        ] at (boxA.north)
        {A: $d=3$};
        \node[
            anchor=south,
            font=\normalfont\bfseries,
            yshift=0.12cm
        ] at (boxB.north)
        {B: $d=4$};
    \end{tikzpicture}%
    }
        \caption{\textbf{Emergence of chimera states induced by clique projection.}
        Comparison of the spatiotemporal dynamics on hypergraphs and their corresponding clique-projected networks for: (A) three-body $d=3$, and (B) four-body $d=4$ interactions, at the parameter sets marked by the bold plus ($\bm{+}$) and lozenge ($\blacklozenge$) symbols in Fig.~\ref{fig:SI_map1}, respectively. The coupling parameters are
        $(\sigma_e,\sigma_c)=(-0.066,0.408)$ for panel A and $(\sigma_e,\sigma_c)=(-0.025,0.18)$ for panel B. In each panel, (a,b) illustrate the spatiotemporal evolution of the activator variable $x_i$, whereas (c,d) display the corresponding spatial snapshots at $t=1480$. (a,c) correspond to the original hypergraph, while (b,d) represent its clique-projected counterpart. Remarkably, for both interaction orders, the original hypergraphs [A(a,c) and B(a,c)]
        exhibit fully incoherent dynamics, whereas their clique-projected counterparts [A(b,d) and B(b,d)] develop well-defined chimera states characterized by the coexistence of coherent and incoherent domains.
        These results reveal that clique projection can promote the emergence of chimera states in parameter regimes where genuine higher-order interactions lead to complete incoherence, highlighting a qualitative dynamical discrepancy between the two representations.
        }
    \label{fig:5}
\end{figure*}

An even more striking scenario is illustrated in Fig.~\ref{fig:5}, where specific combinations of the coupling parameters $(\sigma_e,\sigma_c)$ lead the clique-projected networks to exhibit well-defined chimera states, characterized by the coexistence of distinct coherent and incoherent subpopulation of neurons. This behavior is observed for both three-body interactions [Fig.~\ref{fig:5}A(b,d)] at $(\sigma_e,\sigma_c)=(-0.066,0.408)$ and four-body interactions [Fig.~\ref{fig:5}B(b,d)] at $(\sigma_e,\sigma_c)=(-0.025,0.18)$.
By contrast, the corresponding higher-order interactions on hypergraphs display fully incoherent dynamics, as shown in Figs.~\ref{fig:5}(a,c) and \ref{fig:5}(a,c), respectively. These observations may reveal that, replacing the higher-order interactions with their corresponding pairwise counterparts does not necessarily suppress the emergence of chimera states when attractive and repulsive coupling are considered. Instead, clique projection can qualitatively reorganize the collective dynamics, promoting chimera even in parameter regimes where the original higher-order system remains completely incoherent. Therefore, these findings extend previous studies
\cite{muolo2024phase,muolo2025pinning,kundu2022higher}, which highlighted the important role of higher-order interactions in promoting and sustaining chimera states. In particular, their results showed that chimera states may be substantially weakened or suppressed when higher-order interactions are replaced by their clique-projected counterparts. However, our results reveal a complementary and, in certain regimes, opposite scenario, demonstrating that clique-projected networks can sustain chimera states even when their higher-order counterparts exhibit fully incoherent dynamics [Fig.~\ref{fig:5}]. Thus, this finding suggest that the emergence of chimera states is not not only restricted to the order and topology of the interactions, but also depends on the specific functional form of the coupling and the competition between its attractive and repulsive components.

\section{Conclusion}
\label{sec:conclusion}
In this paper, we have proposed a generalized mathematical framework for chemical and electrical coupling in neuronal hypergraphs that explicitly accounts for higher-order interactions and the interplay between attractive chemical and repulsive electrical coupling, two key mechanisms underlying the emergence collective dynamics in neuronal networks.  The proposed framework has been applied to networks of FitzHugh--Nagumo neurons for different interaction orders or hyperedge sizes, namely $d=2$, $d=3$, and $d=4$, corresponding to three-, four-, and five-body interactions, respectively. Although demonstrated here using the FitzHugh--Nagumo model, the framework is sufficiently general to be extended to other neuronal models.  Through extensive numerical simulations of hypergraph-based neuronal networks with higher-order interactions and their corresponding clique-projected networks with pairwise interactions, we have systematically investigated the combined effects of attractive and repulsive coupling strengths and hyperedge sizes on the emergence and persistence of chimera states. Our results reveal a rich diversity of collective behaviors on both hypergraphs and clique projected networks, ranging from coherent and incoherent dynamics to chimera states. While higher-order interactions
frequently promote broader regions of chimera existence, we have also identified parameter regions of attractive and repulsive coupling strength in which hypergraphs and their clique projections exhibit remarkably similar spatiotemporal patterns, i.e., coherence, incoherence, and chimera states. More strikingly, our analysis shows that in some scenario, chimera states can emerge in clique-projected networks, whereas their genuine higher-order counterparts remain fully incoherent. These observations demonstrate that replacing many-body interactions with effective pairwise couplings does not necessarily suppress chimera formation, but can instead qualitatively reorganize the collective dynamics depending also the coupling functions.
Our findings extend previous studies on chimera states in higher-order
networks by demonstrating that their emergence cannot be attributed solely to interaction order or network topology. Rather, this investigation suggests that it results from a subtle interplay between the mathematical structure of the coupling functions and the competition between attractive and repulsive interactions. In particular, our results show  that clique projection may preserve, suppress, or even promote chimera states depending on the coupling configuration and the explored parameter regime. 

\textit{\textbf{Acknowledgments.}}
ELFN  acknowledges financial support from CAPES (Grant No.~0001). The authors FFF and TN gratefully acknowledges the financial support by National Institute of Science and Technology in Innovative Research in Health Sciences – from Nanotechnology to Artificial Intelligence (INCT PICS) sponsored by Brazil’s National Council for Scientific and Technological Development (CNPq), grant no. 408417/2024-2 and São Paulo Research Foundation (FAPESP), grant no. 2025/26818-7.Also, FFF and TN acknowledge the financial support by FAPESP grant no. 2025/18142-3, FFF thanks CNPq grants no.311989/2025-0. T.N. acknowledges support from the Educating Future Citizens (EFC) project (Project No. SSU2024-00002), funded by the European Union–NextGenerationEU under the Italian National Recovery and Resilience Plan (PNRR).

\bibliography{References}
\end{document}